\documentstyle[amssymb,aps,prb,epsf]{revtex}
\begin{document}
% for two column  activate the line below...                
%\twocolumn[\hsize\textwidth\columnwidth\hsize\csname@twocolumnfalse\endcsname
\title{Josephson Coupling, Phase Correlations, and Josephson Plasma
Resonance in Vortex Liquid Phase. }
\author{A.\ E.\ Koshelev$^a$, L.\ N.\ Bulaevskii$^b$ and M.\ P.\ Maley$^b$} 
\address{$^a$ Materials Science Division,\\ Argonne National Laboratory, \\
Argonne, Illinois 60439 \\
$^b$ Los Alamos National Laboratory, Los Alamos, NM 87545\\
}
\date{\today}
\maketitle

\begin{abstract}
Josephson plasma resonance has been introduced recently as a powerful
tool to probe interlayer Josephson coupling in different regions of
the vortex phase diagram in layered superconductors.  In the liquid
phase, the high temperature expansion with respect to the Josephson
coupling connects the Josephson plasma frequency with the phase
correlation function.  This function, in turn, is directly related to
the pair distribution function of the liquid.  We develop a recipe to
extract the phase and density correlation functions from the
dependencies of the plasma resonance frequency $\omega_p({\bf B})$ and
the $c$ axis conductivity $\sigma_c({\bf B}) $ on the {\it ab}-component
of the magnetic field at fixed {\it c} -component.  Using Langevin
dynamic simulations of two-dimensional vortex arrays we calculate
density and phase correlation functions at different temperatures. 
Calculated phase correlations describe very well the experimental 
angular dependence of the plasma resonance field.  We also demonstrate
that in the case of weak damping in the liquid phase, broadening of
the JPR line is caused mainly by random Josephson coupling arising
from the density fluctuations of pancake vortices.  In this case the
JPR line has a universal shape, which is determined only by
parameters of the superconductors and temperature.
\end{abstract}


%LA-UR-95-2706 
% for two column  activate the line below... 
%\vskip.2pc]

%\narrowtext

\section{Introduction}

Recent studies of Josephson plasma resonance (JPR) absorption in
highly anisotropic layered high-temperature and organic
superconductors \cite
{oph,MatsPRL95,MatsPRL97,MatsPRB97,ShibPRB97,KadowPRB97,%
KakeyaPRB98,OngPhysC97,SakamotoPRB96,MaedaPhysC97}
have proved that this tool provides unique information on Josephson
coupling of layers and on the structure of vortex phases in the
presence of an applied magnetic field.  JPR in layered superconductors
was observed by applying a microwave electric field oriented along the
{\it c} axis which excites charge oscillations between layers.  In the
superconducting state interlayer charge transfer is provided by
tunneling of the Cooper pairs.  In the absence of a magnetic field the
JPR frequency, $\omega _{p}$, at zero temperature is determined by
Josephson interlayer coupling \cite{mish,art93,tach}:
\begin{equation}
\omega _{p}^{2}(B=0)\equiv \omega _{0}^{2}=\frac{8\pi ^{2}cs}{\epsilon
_{c}\Phi _{0}}J_{0}=\frac{c^{2}}{\epsilon _{c}\lambda _{c}^{2}}= \frac{ c^{2}
}{\epsilon _{c}\gamma ^{2}\lambda _{ab}^{2}}  \label{omega0}
\end{equation}
where $J_{0}=c\Phi _{0}/(8\pi ^{2}s\lambda _{c}^{2})$ is the
Josephson current, $\epsilon _{c}$ is the high frequency dielectric
constant for electric fields along the {\it c} axis, $\lambda _{ab}$
and $\lambda _{c}$ are the London penetration lengths for supercurrent
in the $ab$ plane and in the $c$ direction respectively, $\gamma
=\lambda _{c}/\lambda _{ab}$ is the anisotropy ratio and $s$ is the
interlayer spacing.  For the most investigated HTS compound
Bi$_{2}$Sr$_{2}$CaCu$_{2}$O$_{x}$ (Bi-2212) the JPR frequency $\nu
_{p}=\omega _{p}/2\pi $ at low temperatures is in the range of several
hundred gigahertz.  At nonzero temperature, in the presence of
thermally excited quasiparticles, Eq.\ (\ref{omega0}) is valid as
well, if the JPR frequency is well below the scattering rate of
quasiparticles, because under this condition only the Cooper pairs
contribute to plasma oscillations \cite{PokrJS95,artb}.  Then, the JPR
frequency drops with temperature proportional to $1/\lambda
_{c}^{2}(T)$.  This allows one to observe resonance absorption as a
function of temperature at a fixed frequency of uniform microwave
electric field oriented along the {\it c} axis \cite{yuji}.  Recently the 
JPR frequency $\omega_0(T)$ was also measured by sweeping frequency
\cite {gai} and the value $\omega_0(0)/2\pi=125$ GHz was obtained for
slightly underdoped Bi-2212. The same JPR frequency was recently
obtained for a Bi-2212 whisker \cite{matla}. Transport measurements 
\cite{matla} on the mesa 
fabricated from this wisker gave  the critical current density
$J_0(0)=1200$ A/cm$^2$.  This allows one to extract the dielectric
constant $\epsilon_c$ from Eq.~(\ref{omega0}), $\epsilon_c\approx 12$.

A DC magnetic field applied along the {\it c} axis, penetrates inside the 
superconductor in the form of pancake vortices
\cite{BuzFei,ArtKrug,clem}.  In an ideal material at low temperatures
pancakes would form straight lines and have no influence on interlayer
coupling.  Due to very weak coupling, pinning and thermal fluctuations
easily misalign pancakes in different layers, which leads to
suppression of effective Josephson coupling \cite{gkPRB91} and, as a
consequence, to reduction of the Josephson plasma resonance frequency
\cite{bmt,bpmPRL96}.
%the JPR resonance is 
%affected strongly by pancake vortices if their positions in neighboring layers 
%deviate from straight lines parallel to the {\it c} axis \cite{bmt}.  
In real materials pancakes in neighboring layers are always misaligned due
to pinning in the vortex glass phase (below the irreversibility line) and
due to thermal disorder in the vortex liquid phase. Misaligned pancakes
induce a gauge-invariant phase difference $\varphi_{n,n+1}({\bf r})$ between
neighboring layers $n$ and $n+1$ which causes suppression of the JPR
frequency according to the relation \cite{bmt,bpmPRL96}:
\begin{equation}
\omega_p^2(B,T)=\omega_0^2(T){\cal C}(B,T),\; \; {\cal C}(B,T)\equiv\langle
\cos\varphi_{n,n+1}({\bf r})\rangle,  \label{OmCos}
\end{equation}
where $\langle \ldots \rangle$ means thermal average and average over
disorder caused by pinning centers.

The JPR frequency is given by the simple expression (\ref{OmCos}) and is not
sensitive to phase dynamics only if $\omega_p(B,T)$ is much higher than the
characteristic frequencies $\omega _{ps}$ of interlayer phase slips 
induced by vortex motion \cite{kosh}. 
This frequency can be estimated from the {\it ab}
component of static resistivity \cite{KoshPRL96A}, $\omega _{ps}\approx (2\div
4)\cdot 10^8$[1/s]$\cdot T$[K]$\cdot \rho _{ab}$. 
We will show that, more directly, $\omega _{ps}$ can be extracted from both
the {\it c} axis conductivity and plasma frequency. These estimates show
that typically $\omega_{ps}$ in the liquid region varies from $10^{-4}\omega
_{p}$ at $T\approx 40$ K to $10^{-2}\omega _{p}$ at $T\approx 70$ K. Thus,
the plasma frequency is determined by a snapshot of vortex positions. For
this reason JPR study provides information on the equilibrium properties of
the vortex lattice though it is dynamic in nature.

The plasma frequency is mainly sensitive to the average value of $\cos
\varphi _{n,n+1}({\bf r})$ even if its local variations are very large,
provided the random part of $\cos \varphi _{n,n+1}({\bf r})$ changes
rapidly in space.  In this case these rapid variations are averaged out
by the smooth external oscillating electric field.  Space variations of
$\cos \varphi _{n,n+1}({\bf r })$ result in mixing of phase collective
modes with different momenta and frequency.  Such a mixing is
responsible for resonance line broadening\cite{bdmb,LineShapePRB99}. 
Mixing is determined by Josephson coupling and is proportional to $
l_{\varphi }^{2}/\lambda _{J}^{2}$, where $\lambda _{J}=\gamma s$ is the
Josephson length, and $l_{\varphi }$ is the characteristic length of the
phase correlation function
\begin{equation}
S({\bf r},T,{\bf B})=\langle \cos [\varphi _{n,n+1}({\bf r})-\varphi
_{n,n+1}(0)]\rangle .  \label{coscorr}
\end{equation}
Averaging in space is effective for $l_{\varphi }\ll \lambda _{J}$. 
This condition is fulfilled for fields $B\gg B_{J}=\Phi _{0}/\lambda
_{J}^{2}$ in a vortex phase with strong disorder along the {\it c}
axis, when $l_{\varphi }\approx a$.  Such a situation is realized in
the vortex liquid phase where interlayer correlations of vortex
positions are very weak and also in the Bean critical state obtained
by sweeping magnetic field after zero field cooling.  $l_{\varphi }$
is of the order of $a$ also in the vortex glass phase obtained by
cooling in high magnetic fields above the ``second peak'' field
($\approx 300-500$ G), where the three-dimensional vortex lattice is
strongly disordered according to $\mu ^{+}$SR \cite{muSR} and neutron
scattering \cite{cubbit} measurements. In this decoupled phase, interaction 
of pancake vortices in different layers is very weak, see 
discussion in Ref.~\onlinecite{kv}.

Therefore, the key parameter which determines field and temperature
dependence of the JPR frequency is ${\cal C}(B,T)$.  This allows the 
use of JPR as a sensitive probe of the vortex state.  The field dependence of
$\omega_p$ allows one to observe JPR absorption as a function of
magnetic field by sweeping the magnetic field $B$ at fixed frequency
$\omega$ of applied microwave electric field.  In this way the
resonance magnetic field $B_r$ was obtained as a function of
temperature and frequency $\omega$, $\omega_{p}(B_{r},T)=\omega$, see,
e.\ g., Refs.~\onlinecite{oph,MatsPRL95,ShibPRB97,KadowPRB97}.  
Tsui {\it et al.}
\cite{oph} found the field and temperature dependence of $\omega_p$ in
the vortex state of Bi-2212 by measuring $B_r(\omega,T)$ at several
frequencies and using field sweeping (FS) after zero field cooling. 
They found that the resonance frequency depends nonmonotonically on
temperature at fixed field and has a maximum at the irreversibility
line.  Measuring $B_r(\omega,T)$ in the field cooling (FC) and FS
modes, Matsuda {\it et al.} \cite{MatsPRL95} demonstrated that the plasma
frequency is history dependent below the irreversibility line.  
Recenly the Josephson plasma resonance has also been measured in the vortex
crystal state at small fields\cite{ShibPRL99,GaifPRL00}. Field dependence
of the JPR frequency in this phase is well described by the model based
on the suppression of the Josephson coupling due to elastic fluctuations of
pancake vortices\cite{JPRcryst}.

In the vortex liquid phase, for the magnetic field $B_{z}$ oriented along
the {\it c} axis, the field and temperature dependence of JPR 
frequency is approximately described as \cite{oph,MatsPRL95,MaedaPhysC97}
\begin{equation}
\omega _{p}^{2}(B_{z},T)\propto \omega _{0}^{2}(T)B_{z}^{-\nu }T^{-1}
\label{lf}
\end{equation}
with $\nu $ being slightly smaller than unity ($\approx 0.8$).  

The influence of vortices on interlayer coupling in the vortex liquid
was also probed by direct {\it c}-axis transport measurements
\cite{Rodriguez,cho,yas,yurgens,suzuki}.  In particular, pancake
misalignment suppresses the Josephson current $J_{c}(T,B_{z})$ as
compared to its zero field value $J_{0}(T)$, as $J_{c}(T,B_{z})\approx
J_{0}(T){\cal C}(T,B_{z})$, so that $J_{c}(T,B_{z})$ is proportional
to $\omega _{p}^{2}(B_{z},T)$.  The field dependence
$J_{c}(T,B_{z})\propto B_{z}^{-1}$, similar to Eq.\ ( \ref{lf}) was
indeed found for the {\it c}-axis critical current
\cite{yas,yurgens,suzuki}.

Behavior of JPR in the liquid phase described by Eq.~(\ref{lf}) was
explained in Ref.~\onlinecite{kosh} in the framework of the high
temperature expansion with respect to Josephson coupling.
%, based on the relation (\ref{OmCos}).  
% In this approach magnetic coupling of vortices in different layers is
% neglected and perturbation theory for Josephson coupling is used
% resulting in:
This approach provides a general relation, connecting the average
cosine factor ${\cal C}(B,T)$ with the phase correlation function
$S({\bf r},T,{\bf B})$ defined  by Eq.\ (\ref{coscorr}),
\begin{equation}
{\cal C}\approx \frac{E_{J}}{2T}\int d{\bf r}S({\bf r},T,B_{z})=f\frac{
E_{0}B_{J}}{2TB_{z}},  
\label{cfun}
\end{equation}
where thermal average $\langle \ldots \rangle$ in Eq.\ (\ref{coscorr})
is now done without the Josephson coupling, $E_{J}=E_{0}/\lambda
_{J}^{2}$ is the energy of Josephson interlayer coupling per unit
area, $E_{0}=\Phi _{0}^{2}s/16\pi ^{3}\lambda _{ab}^{2}$, and
\begin{equation}
f=\frac{B}{\Phi _{0}}\int d{\bf r}S({\bf r},T,B_{z})  
\label{fexpr}
\end{equation}
is a dimensionless universal function of order unity.  The
relation (\ref {cfun}) was derived assuming that: a) $\langle \cos
[\varphi _{n,n+1}({\bf r} )]\rangle =0$ if the Josephson coupling is
absent, and b) $S({\bf r},T,B_{z})$ is determined by vortices only,
and thus its correlation length $l_{\varphi }\approx a$.  Such a
perturbative approach is valid if the Josephson energy in the area
$l_{\varphi }^{2}$ is small compared with $T$, i.e. $ E_{0}l_{\varphi
}^{2}/\lambda _{J}^{2}\approx E_{0}a^{2}/\lambda _{J}^{2}\ll T$.  This
condition is fulfilled in magnetic fields $B_{z}\gg B_{J}(E_{0}/T)$
and, under such condition, one gets $\omega _{p}(B,T)\ll \omega
_{0}(T)$.  For Bi-2212 crystals with $\gamma $ of the order of several
hundred the high temperature expansion approach is valid practically
at all fields above $\approx 100$ G. 

The same physical picture provides prediction for dependence of the JPR
frequency on the in-plane field $B_{x}$ .  In the decoupled phase at
$B_{z}\gg B_{J}$ the in-plane field produces an additional linearly
growing term in the phase difference, $2\pi B_{x}ys/\Phi_{0}$, which
accounts for the phase difference due to the vector potential
corresponding to $B_{x}$.  Therefore, the dependence of $\omega_{p}^{2}$
on $B_{x}$ is simply determined by the Fourier transform of the phase
correlation function $S({\bf r},T,B_{z})$.  It was also assumed in Ref.\
\onlinecite{kosh} that $S({\bf r},T,B_{z})$ can be approximated by a
Gaussian, $S({\bf r})=\exp (-\pi B_zr^{2}/2f\Phi _{0})$.  In this case
the dependence of JPR on the in-plane field can be presented in the
analytical form:
\begin{equation}
\omega _{p}^{2}(B_{x},B_{z},T)\approx \omega _{p}^{2}(B_{z},T)\exp
(-f\pi s^{2}B_{x}^{2}/\Phi _{0}B_{z}).  \label{alf}
\end{equation}
The exponential factor is the Fourier component of $S({\bf r})$.  Such
dependence was found to describe fairly well the experimental angular
dependencies of the resonance field \cite{MatsPRB97,OngPhysC97}.

In this paper we develop a detailed theory of the Josephson plasma
resonance in the liquid phase based on perturbation theory with
respect to the Josephson coupling.  This approach is justified because
$\omega _{p}^{2}/\omega _{0}^{2}={\cal C}(B,T)$ is well below unity in
this state.  We will extend the analysis of Ref.~\onlinecite{kosh} by
taking into account spin-wave type phase fluctuations in addition to
the effect of vortices on JPR. In this approach $S({\bf
r},T,B)=S_{r}({\bf r} ,T)S_{v}({\bf r},T,B)$, where $S_{r}( {\bf
r},T)$ is determined by spin-wave type phase fluctuations, while
$S_{v}( {\bf r},T,B)$ is the phase correlation function induced by
vortices.  We will calculate $S_{r}({\bf r},T)$ and express
$S_{v}({\bf r},T,B)$ via the correlation function of vortex density in
the system without Josephson coupling.  We show that, at temperatures
above the melting point, the coupling between two dimensional pancake
liquids in different layers is very weak and can be neglected.  If, in
addition, pinning inside the layers is neglected, then the density
correlation function coincides with the density correlation function
of the two-dimensional one-component Coulomb plasma \cite{caillol}. 
Accurate analysis shows that the Gaussian dependence of $S_{v}({\bf
r})$ used in Refs.\ \onlinecite{kosh} and \onlinecite{MatsPRB97} actually
exists only at $r \lesssim a$; at large distances $S_{v}({\bf r})$
decays exponentially, $S_{v}({\bf r})\propto \exp (-r/l)$.  This means
that the dependence of JPR frequency on the in-plane field $B_{x}$ is also
not exactly Gaussian, and is given by the Fourier transform of the
phase correlation function, $f_{s}(b)\approx (1/a^{2})\int d{\bf r} 
\exp (i bx/a) S({\bf r})$,
\begin{equation}
\omega _{p}^{2}(B_{x},B_{z})=\omega _{p}^{2}(0,B_{z})f_{s}(B_{x}/\sqrt{B_{z}H_{0}}),
\label{5}
\end{equation}
with $H_{0}=\Phi _{0}/(2\pi s)^{2}$.  As the function $f_{s}(b)$ is
directly connected to the phase and density correlation functions,
this equation allows us to obtain information on the structure of the
vortex state from experimental study of JPR. We calculated numerically
the two-dimensional density correlation functions using Langevin
dynamics simulations and used these functions to calculate the phase
correlation functions $S_{v}({\bf r} ,T,B)$ at different temperatures. 
Influence of weak interlayer coupling and pinning is investigated
perturbatively.  We found that the function $f_{s}(b)$ in Eq.\ (\ref{5})
calculated from numerical data gives a better description of the angular
dependence of the resonance field than the naive Gaussian ansatz
(\ref{alf}). Part of these results has been published in a short paper
\cite{KoshBulMay:PRL98}.

We will investigate also the shape of the resonance line.  We
demonstrate that in the case of weak damping in the liquid phase the
plasma resonance line acquires a universal shape due to
fluctuation-induced inhomogeneities of the interlayer phase difference. 
This mechanism gives a natural explanation for experimentally observed
asymmetric broadening of the line.  The tail at high frequencies arises
due to mixing of propagating plasma modes by random Josephson coupling. 
The low frequency tail is caused by plasma modes localized in the rare
large areas where the Josephson coupling is suppressed due to
fluctuations, similar to the well known Lifshitz tail in the electron
density of states in disordered semiconductors.  These results were
published in the short communication \onlinecite{LineShapePRB99}.  In
this paper we present their detailed derivations.  At high frequences
(fields) we calculate the shape of the line using perturbative expansion
with respect to inhomogeneous Josephson coupling.  At frequencies below
the resonance frequency we use the method of optimal fluctuation to
calculate the exponential tail of the JPR line.

The paper is organized as follows.  In Section II we present general
thermodynamic relations for a layered superconductor in the mixed state
and discuss phase dynamics.  Section III is devoted to Josephson
coupling in the liquid state.  We obtain the high temperature expansion
for the parameter ${\cal C }(B,T)$, derive general relations connecting
the JPR frequency with the pair distribution functions, and calculate
the pair distribution functions numerically.  We also establish a
relation between the plasma frequency and the static c-axis
conductivity.  In Section IV we calculate the shape of the JPR line in
the liquid state due to the inhomogeneous broadening mechanism.

\section{General formalism}

\subsection{Thermodynamics}

We consider a layered superconductor in the vortex state. In the most part of
the field-temperature phase diagram below $T_{c}$ the 
relevant degrees of freedom
are the vortex coordinates ${\bf r}_{n\nu }$ and phase spin-wave type
degrees of freedom $\varphi _{n,n+1}^{(r)}({\bf r})$. The energy functional
in terms of these degrees of freedom can be written as \cite{bdmb}: 
\begin{equation}
{\cal F}\{{\bf r}_{n\nu },\varphi _{n,n+1}^{(r)}({\bf r})\}={\cal F}_{em}( 
{\bf r}_{n\nu })+{\cal F}_{pin}({\bf r}_{n\nu })+{\cal F}_{\varphi }\{{\bf r}
_{n\nu },\varphi _{n,n+1}^{(r)}({\bf r})\}.  \label{mainf}
\end{equation}
Here ${\cal F}_{em}({\bf r}_{n\nu })$ is the functional which accounts for
the two-dimensional energy of pancakes and their electromagnetic interaction
in different layers \cite{BuzFei,ArtKrug,clem}: 
\begin{equation}
{\cal F}_{em}({\bf r}_{n\nu })=\frac{E_{0}}{4\pi }\int d{\bf k}
dq\sum_{n,m,\nu ,\nu ^{\prime }}\frac{\exp [i{\bf k}\cdot ({\bf r}_{n\nu }- 
{\bf r}_{m\nu ^{\prime }})+iq(n-m)]}{k^{2}[1+\lambda
_{ab}^{-2}(k^{2}+Q^{2})^{-1}]},  \label{f2}
\end{equation}
where $Q^{2}=2(1-\cos q)/s^{2}$. The contribution ${\cal F}_{pin}({\bf r}
_{n\nu })$ accounts for pinning: 
\begin{equation}
{\cal F}_{pin}({\bf r}_{n\nu })=\sum_{n,\nu }\int d{\bf r}V_{pin}({\bf r}
)\delta ({\bf r}-{\bf r}_{n\nu }),  \label{f4}
\end{equation}
where $V_{pin}({\bf r})$ is the pinning potential for vortices, see, e.g.,
Ref.~ \onlinecite{Blatter}. The last term is the energy of intralayer
currents associated with spin-wave type of excitations and the Josephson
energy. It depends on the regular part of the phase difference and vortex
coordinates: 
\begin{eqnarray}
&&{\cal F}_{\varphi }\{{\bf r}_{n\nu },\varphi _{n,n+1}^{(r)}\}=
\label{vfunc} \\
&&E_{0}\int d{\bf r}\left\{ \frac{1}{2}\sum_{mn}L_{nm}{\bf \nabla} \varphi
_{n,n+1}^{(r)}\cdot {\bf \nabla}\varphi _{m,m+1}^{(r)}+\frac{1}{\lambda _{J}^{2}}
\sum_{n}\left[ 1-\cos (\varphi _{n,n+1}^{(v)}+\varphi _{n,n+1}^{(r)})\right]
\right\}.  \nonumber
\end{eqnarray}
Here ${\bf \nabla}=(\partial/\partial x,\partial/\partial y)$. The mutual
inductance of layers, $L_{nm}$, is: 
\begin{equation}
L_{nm}=\int_{-\pi }^{\pi }\frac{dq}{2\pi }L(q)\cos (n-m)q \approx \frac{
\lambda_{ab}}{2s}\exp\left[-\frac{|n-m|s}{\lambda_{ab}} \right],\ \ \ L(q)=
\frac{1 }{2(1-\cos q)+s^{2}/\lambda _{ab}^{2}}.
\end{equation}
The phase $\varphi _{n,n+1}^{(v)}({\bf r})$ is the singular part of the
phase difference induced by vortices at positions ${\bf r}_{n\nu }$ and
${\bf r}_{n+1\nu }$ when Josephson coupling is absent ($1/\lambda
_{J}=0$):
\begin{equation}
\varphi _{n,n+1}^{(v)}({\bf r})=\sum_{\nu }[\phi _{v}({\bf r}
-{\bf r}_{n\nu })-\phi _{v}({\bf r}-{\bf r}_{n+1,\nu })],  \label{2D}
\end{equation}
where $\phi _{v}({\bf r})\equiv \arctan(y/x)$ is the polar angle of the
point ${\bf r}$.

The average cosine of the total phase difference is given by the expression: 
\begin{eqnarray}
&&{\cal C}=1+\frac{T}{NS}\frac{\partial }{\partial E_{J}}\langle \ln
Z\rangle _{dis}, \\
&&Z=\int D\varphi _{n,n+1}^{(r)}({\bf r})D{\bf r}_{n\nu }\exp \left[ -{\cal 
F }\{{\bf r}_{n\nu },\varphi _{n,n+1}^{(r)}({\bf r})\}/T\right] .
\end{eqnarray}
Here $\langle \ldots \rangle _{dis}$ is an average over disorder 
(pinning), $N$ is the total number of layers, and $S$ is the total area 
of the layer. 

\subsection{Phase dynamics}

Time variations of the phase difference leading to excitations of
quasiparticles and the deviations of the quasiparticle distribution function
from equilibrium should be accounted for in the time-dependent equation for
the phase difference. General equations describing phase dynamics are still
under discussion and they depend on the mechanism of pairing, see Refs.~
\onlinecite{ArtKobPRL97,koy,Ryn}. At the phenomenological level, the dynamic equations for
the phase difference $\varphi _{n,n+1}({\bf r},t)$ can be derived as
follows. Introducing the in-plane superfluid momentum ${\bf p}
_{n}=\nabla\phi _{n}- (2\pi/\Phi _{0}){\bf A}_{n}$, we can express the
in-plane magnetic field as 
\[
{\bf H}_{\perp }=\frac{\Phi _{0}}{2\pi s}{\bf n}_{z}\times \left( \nabla
\varphi _{n,n+1}-s\nabla _{z}{\bf p}_{n}\right) 
\]
where ${\bf n}_{z}$ is the unit vector in z-direction, $\nabla _{z}{\bf p}
_{n}\equiv ({\bf p}_{n+1}-{\bf p}_{n})/s$. Using this representation the
z-component of the Maxwell equation ${\rm curl} {\bf B}=(4\pi/c) {\bf j}
+(1/c)\partial {\bf D}/\partial t$ can be written as 
\begin{equation}
\frac{4\pi }{c}j_{z}{\bf +}\frac{\varepsilon _{c}}{c}\frac{\partial E_{z}}{
\partial t}=\frac{\Phi _{0}}{2\pi s}\left( \nabla ^{2}\varphi
_{n,n+1}-s\nabla _{z}\nabla{\bf p}_{n}\right)  \label{z-compMaxw}
\end{equation}
where the current $j_{z}$ consists of the Josephson and quasiparticle
contributions, $j_{z}=J_{0}\sin\varphi _{n,n+1}+j_{{\rm q}z}$. From the
in-plane component of the Maxwell equation we obtain the independent
equation for $\nabla{\bf p}_{n}$ 
\begin{equation}
\left( -\nabla _{z}^{2}+\lambda_{ab} ^{-2}\right) \nabla {\bf p}_{n}=- \frac{
8\pi ^{2}}{c\Phi _{0}}\nabla\left( {\bf j}_ {n}+ \frac{\varepsilon _{ab}}{
4\pi }\frac{\partial {\bf E}_{\perp }}{\partial t} \right) -\frac{1}{s}
\nabla _{z}\nabla^{2}\varphi _{n-1,n},  \label{divp}
\end{equation}
Here $\epsilon_{ab}$ is the high frequency dielectric constant for the
in-plane electric field. We split the in-plane current into the
quasiparticle and superconducting components, ${\bf j}={\bf j}_{{\rm s}}+
{\bf j}_{{\rm q}} $ with ${\bf j}_{{\rm s}}=(c\Phi _{0}/8\pi
^{2}\lambda_{ab} ^{2}){\bf p}_{n}$. In the following we consider the case
when the in-plane equilibration occurs at time scales much shorter than the
inverse plasma frequency. This allows us to neglect the in-plane
quasiparticle current and electric field in Eq.\ (\ref{divp}). Solving
formally Eq.\ (\ref{divp}) with respect to $\nabla {\bf p}_{n}$ and
substituting it into Eq.\ (\ref{z-compMaxw}) we obtain 
\begin{equation}
\sin \varphi _{n,n+1} -\lambda _{J}^{2}\sum_{m}L_{nm}\nabla ^{2}\varphi
_{m,m+1}+\frac{j_{{\rm q}z}}{J_{0}}{\bf +}\frac{\varepsilon _{c} }{4\pi J_{0}
}\frac{\partial E_{z}}{\partial t}=0  \label{PhaseDyn1}
\end{equation}
To obtain a closed set of equations for $\varphi _{n,n+1}$, we have to
connect $E_{z}$ with $\partial \varphi _{n,n+1}/\partial t$. In a common
Josephson junction these quantities are connected by the famous Josephson
relation $\partial \phi _{n,n+1}/\partial t=(2\pi cs/\Phi _{0})E_{z}$.
However, it was shown\cite{art93,koy} that in atomically layered
superconductors this simple relation is violated due to the charging of
layers. The charging effect leads to the variations of the chemical
potential $\mu _{n}$ in the layer $n$, and, in general, the time derivative
of $\varphi _{n,n+1}$ can be written as 
\begin{equation}
\frac{\partial \varphi _{n,n+1}}{\partial t}=\frac{2\pi cs}{\Phi _{0}}\left(
E_{z}-\nabla _{z}\frac{\mu _{n}}{e}\right) .  \label{ModifJos}
\end{equation}
where $\mu _{n}$ is related with the variation of the charge $\rho _{n}$ as 
\begin{equation}
\mu _{n}=\frac{4\pi e\lambda _{D}^{2}}{\varepsilon _{c}s}\rho _{n}
\label{mu-rho}
\end{equation}
where $\,\lambda _{D}$ is the Debye screening length. 
%estimated as $\theta s$ with $\theta<1$.  
The charge density is related to electric field by the Poisson equation 
\begin{equation}
\nabla _{z}E_{z}=\frac{4\pi }{\varepsilon _{c}}\rho _{n}  \label{Poisson}
\end{equation}
Combining Eqs.(\ref{ModifJos}), (\ref{mu-rho}), and (\ref{Poisson}) we
obtain relation between $E_{z}$ and $\partial \varphi _{n,n+1}/\partial t$ 
\begin{equation}
\frac{\partial \varphi _{n,n+1}}{\partial t}=\frac{2\pi cs}{\Phi _{0}}\left(
1-\lambda _{D}^{2}\nabla _{z}^{2}\right) E_{z}.  \label{phi-E}
\end{equation}
Finally, the quasiparticle current is determined by the quasiparticle
conductivity $\sigma _{{\rm q}z}$ 
\begin{equation}
j_{{\rm q}z}=\frac{\Phi _{0}\sigma _{{\rm q}z}}{2\pi cs}\frac{\partial
\varphi _{n,n+1} }{\partial t}  \label{normj}
\end{equation}
The interlayer quasiparticle conductivity in superconductors with d-wave
pairing remains nonzero even at $T\rightarrow 0$ and it was found to be
$\approx 1-2$ (kohm cm)$^{-1}$ in Bi-2212 crystals at 4 K and $B=0$
\cite{Lat} depending weakly on $B$ \cite{MorozPRL00}.  Eqs.\
(\ref{PhaseDyn1}), (\ref{phi-E}), and (\ref{normj}) form a closed set of
equations for $\varphi_{n,n+1}$.  They describe phase dynamics when
deviations of the quasiparticle distribution function from that at
equilibrium can be neglected.  Deviations from equilibrium and charging
effects may lead to the broadening of JPR resonance in addition to the
effect of dissipation caused by quasiparticles.  It is unclear at that
moment how important are nonequilibrium effects because until now they
were studied only in the framework of simple microscopic models
\cite{Ryn}, which may be unadequite for HTS.

As we are mainly interested in the effect of pancake vortices on JPR in the
following, we will use simplified equations in which we do not account for
quasiparticle conductivity \cite{zam} and charging effects \cite{art93,koy}.
This simplest form of equation for $\varphi _{n,n+1}^{(r)}({\bf r},t)$ which
describes phase variations at fixed vortex positions is: 
\begin{equation}
\frac{1}{\omega _{0}^{2}}\frac{\partial ^{2}}{\partial t^{2}}\varphi
_{n,n+1}^{(r)}-\lambda _{J}^{2}\sum_{m}L_{nm}\nabla ^{2}\varphi
_{m,m+1}^{(r)}+\sin \left[ \varphi _{n,n+1}^{(r)}+\varphi _{n,n+1}^{(v)}( 
{\bf r})\right] =\frac{1 }{4\pi J_{0}}\frac{\partial }{\partial t}{\cal D}
_{z}(t),
\end{equation}
where ${\cal D}_{z}(t)$ is the homogeneous in space alternating external
electric field, and we assume that $\varphi _{n,n+1}^{(v)}({\bf r})$ is time
independent, because characteristic frequencies of vortex motion are well
below the plasma frequency as discussed above. For small amplitude
variations we expand the phase difference as: 
\begin{equation}
\varphi _{n,n+1}({\bf r},t)=\varphi _{n,n+1}^{(0)}({\bf r})+\varphi
_{n,n+1}^{\prime }({\bf r},t).
\end{equation}
Here $\varphi _{n,n+1}^{(0)}({\bf r})$ is the solution of the stationary
equation. We obtain for the Fourier component of the phase difference
induced by the external electric field, $\varphi _{n,n+1}^{\prime }({\bf r}
,\omega )$, the equation 
\begin{equation}
\frac{\omega ^{2}}{\omega _{0}^{2}}\varphi _{n,n+1}^{\prime }+\lambda
_{J}^{2}\sum_{m}L_{nm}\nabla ^{2}\varphi _{m,m+1}^{\prime }-\varphi
_{n,n+1}^{\prime }\cos [\varphi _{n,n+1}^{(0)}({\bf r})]=-\frac{ i\omega }{
4\pi J_{0}}{\cal D}_{z}(t),  \label{LinPhasDyn}
\end{equation}
The homogeneous part of the averaged electric field between layers $n$ and $
n+1$ is 
\begin{equation}
\langle E_{z}(\omega )\rangle =\frac{i\omega \Phi _{0}}{2\pi cs}\langle
\varphi _{n,n+1}^{\prime }\rangle
\end{equation}
and JPR absorption is determined by the imaginary part of the inverse
dielectric function, $1/\epsilon(\omega)=\langle E_{z}(\omega )\rangle/ 
{\cal D}_z(\omega)$. Expanding $\varphi _{n,n+1}^{\prime }({\bf r})$ in
terms of the eigenfunctions $\Psi_{\alpha n}({\bf r})$ of the
Schr\"{o}dinger-like equation 
\begin{equation}
-\lambda _{J}^{2}\sum_{m}L_{nm}\nabla ^{2}\Psi_{\alpha m}+\cos [\varphi
_{n,n+1}^{(0)}({\bf r})]\Psi_{\alpha n}=\frac{\omega_{\alpha} ^{2}}{\omega
_{0}^{2}}\Psi_{\alpha n},  \label{EigenEq}
\end{equation}
we obtain a useful expression for ${\rm Im}(1/\epsilon(\omega)$ 
\begin{equation}
{\rm Im}\frac{1}{\epsilon(\omega)}=\frac{\pi\omega}{2\epsilon_c} \int d{\bf r
}\left \langle \sum_{\alpha,n}\Psi_{\alpha 0}(0)\Psi^*_{\alpha n}({\bf r}
)\delta(\omega-\omega_{\alpha})\right \rangle,
\end{equation}
where $\langle ...\rangle$ is the average over configurations of $
\cos\varphi_{n,n+1}^{(0)}({\bf r})$. This expression will be used in Sec.\ V
to calculate the shape of the JPR line.

\section{Josephson coupling in the liquid state}

\subsection{High temperature expansion}

In the vortex liquid phase we can use perturbation theory with respect to
Josephson coupling, which relates the cosine of the total phase difference
with the phase correlation function $S({\bf r)}$ as in Eq.~(\ref{cfun}).
Splitting the total phase difference $\varphi _{n,n+1}({\bf r})$ into
vortex, $\varphi _{n,n+1}^{(v)}({\bf r})$, and regular $\varphi
_{n,n+1}^{(r)}({\bf r})$ contributions, we obtain: 
\begin{equation}
\langle \cos [\varphi _{n,n+1}^{(v)}+\varphi _{n,n+1}^{(r)}]\rangle = \frac{
E_{J}}{2T}\int d{\bf r}\langle \cos [\varphi _{n,n+1}({\bf r})-\varphi
_{n,n+1}(0)]\rangle= \frac{E_{J}}{2T}\int d{\bf r}S_{v}({\bf r})S_{r}({\bf r}
),  \label{cosGvGs}
\end{equation}
where the correlation function of phases $S_{v}({\bf r})=\langle \cos
[\varphi _{n,n+1}^{(v)}({\bf r})-\varphi _{n,n+1}^{(v)}(0)]\rangle $ is
determined by vortex positions, 
\begin{eqnarray}
&&S_{v}({\bf r})=\langle Z_{v}^{-1}\int D{\bf r}_{n\nu }\cos [\varphi
_{n,n+1}^{(v)}({\bf r})-\varphi _{n,n+1}^{(v)}(0)]\exp \{-(1/T)[{\cal F}
_{em}+{\cal F}_{pin}]\}\rangle _{dis},  \nonumber \\
&&Z_{v}=\int D{\bf r}_{n\nu }\exp \{-(1/T)[{\cal F}_{em}+{\cal F}_{pin}]\}, 
\nonumber
\end{eqnarray}
and $S_{r}({\bf r})=\langle \cos [\varphi _{n,n+1}^{(r)}({\bf r})-\varphi
_{n,n+1}^{(r)}(0)]\rangle $ describes the effect of spin-wave type phase
fluctuations (phase difference fluctuations in the absence of vortices), 
\begin{eqnarray}
&&S_{r}({\bf r})=Z_{s}^{-1}\int D\varphi _{n,n+1}^{(r)}\cos [\varphi
_{n,n+1}^{(r)}({\bf r})-\varphi _{n,n+1}^{(r)}(0)]\exp
[-(E_{0}/2T)\sum_{nm}L_{nm}\nabla \varphi _{n,n+1}^{(r)}\cdot \nabla \varphi
_{m,m+1}^{(r)}],  \nonumber \\
&&Z_{s}=\int D\varphi _{n,n+1}^{(r)}\exp [-(E_{0}/2T)\sum_{nm}L_{nm}\nabla
\varphi _{n,n+1}^{(r)}\cdot \nabla \varphi _{m,m+1}^{(r)}].  \nonumber
\end{eqnarray}

In the lowest order in $E_J$, spin-wave type of fluctuations and vortex
coordinates degrees of freedom do not interact. In the Gaussian
approximation for spin-wave type of phase fluctuations, we obtain: 
\begin{equation}
S_{r}({\bf r})=\exp \left (-\frac{T}{(2\pi)^3}\int d{\bf k}dq\frac{1-\cos ({\bf k} {\bf r})}{
E_{0}L(q)k^{2}}\right )=\left(\frac{\xi _{ab}}{r}\right)^{2\alpha },\; \;
\alpha =\frac{T}{2\pi E_{0}}.  
\label{scff}
\end{equation}
We cut off the logarithmic divergence in the integration over ${\bf k}$ is at $
k\approx 1/\xi _{ab}$, where $\xi _{ab}$ is the superconducting correlation
length. As we show below, a particular value of this cutoff has no influence
on observable quantities

\subsection{Relation between the phase correlation function $S_{v}({\bf r})$
and the density correlation function of the liquid}

Next we express the phase correlation function $S_{v}({\bf r},T,B)$ via the
density correlation function of pancake vortices induced by the $B_{z}$
component of the magnetic field. Let us consider the phase difference
between points ${\bf r}$ and $0$ in the layer $n$ induced by pancakes with
density $\rho _{n}({\bf R})=\sum_{\nu}\delta({\bf R}-{\bf r}_{n\nu})$.
Denoting the phase in the layer $n$ at the point ${\bf r}$ by $\phi _{n}(
{\bf r})$ we obtain 
\begin{equation}
\phi _{n}({\bf r})-\phi _{n}(0)=\int d{\bf R}\rho _{n}({\bf R})\beta ({\bf r}
,{\bf R}),
\end{equation}
where 
\begin{equation}
\beta ({\bf r},{\bf R})=\phi _{v}({\bf r}/2-{\bf R})-\phi _{v}(-{\bf r}/2- 
{\bf R}) 
\end{equation}
is the angle at which the segment connecting points $-{\bf r}/2$ and ${\bf r}
/2$ is seen from the point ${\bf R}=(X,Y)$. $\beta ({\bf r},{\bf R})$ has a
jump of $2\pi $ when point ${\bf R}$ intersects segment $[-{\bf r}/2,{\bf r}
/2]$. Choosing the $x$ axis along the segment $[-{\bf r}/2,{\bf r}/2]$, 
we can represent $\beta ({\bf r},{\bf R})$  as 
\begin{equation}
\beta ({\bf r},{\bf R})=\arcsin \frac{Yr}{ [(R^{2}+r^{2}/4)^{2}-(Xr)^{2}]^{1/2}},
\end{equation}
For the phase difference $\varphi _{n,n+1}^{(v)}({\bf r})-\varphi
_{n,n+1}^{(v)}(0)$ between layers $n$ and $n+1$ in the presence of both
components $B_{x}$ and $B_{z}$ we obtain 
\begin{eqnarray}
&&\varphi _{n,n+1}^{(v)}({\bf r})-\varphi _{n,n+1}^{(v)}(0)=\Phi _{v}({\bf r}
)+(2\pi /\Phi _{0})\int_{ns}^{(n+1)s}dz\left[ A_{z}({\bf r} ,z)-A_{z}(0,z) 
\right] ,  \label{vp} \\
&&\Phi _{v}({\bf r})=\int d{\bf R}[\rho _{n}({\bf R})-\rho _{n+1}({\bf R}
)]\beta ({\bf r},{\bf R}).  \label{Ph}
\end{eqnarray}
Here $A_{z}(y)=-B_{x}y$ is the vector potential for the $B_{x}$ component of
magnetic field. We assume here that $B_{x}$ is large enough so that it
penetrates almost freely into the sample and therefore has no influence on
the phase fluctuations.

In the following we will use the Gaussian approximation for the random
function $\Phi _{v}({\bf r})$, i.e. we neglect irreducible higher-order
correlation functions. Then 
\begin{equation}
\langle \cos [\Phi _{v}({\bf r})]\rangle \approx \exp [-F_{v}({\bf r})],\ \
\ F_{v}(r)=\langle \lbrack \Phi _{v}({\bf r})]\rangle ^{2}/2.  \label{appr}
\end{equation}
Therefore the vortex-induced phase fluctuations are characterized by the the
correlation function 
\begin{equation}
S_{v}({\bf r})=\exp [-F_{v}(r)]\cos (2\pi sB_{x}y/\Phi _{0}).  \label{GvFv}
\end{equation}
Using Eqs.\ (\ref{Ph}) and (\ref{appr}) we connect $F_{v}(r)$ with the
following density correlation function 
\begin{eqnarray}
K({\bf r}) &=&(1/2)\langle \lbrack \rho _{n}({\bf r})-\rho _{n+1}({\bf r}
)][\rho _{n}(0)-\rho _{n+1}(0)]\rangle  \nonumber\\
&\equiv&K_{0}({\bf r})-K_{1}({\bf r})
\label{DenCorr} 
\end{eqnarray}
by the relation 
\begin{equation}
F_{v}(r)=\int d{\bf R}d{\bf R}_{1}K({\bf R}-{\bf R}_{1})\beta ({\bf r},{\bf R
})\beta ({\bf r},{\bf R}_{1}),  \label{zero}
\end{equation}
where $K_{n}({\bf r})=\left\langle \delta \rho _{0}({\bf r})\delta
\rho _{n}(0)\right\rangle$ and  $\delta \rho _{n}({\bf r})=\rho _{n}({\bf
r})-n_{v}$ is the deviation of vortex density from its average value
$n_{v}=B_{z}/\Phi _{0}$ in the layer $n$ at point ${\bf r}$.  We will
use the representation
\begin{equation}
K({\bf r})=n_{v}\delta ({\bf r})+n_{v}^{2}h(r),\ \ \ n_{v}\int d{\bf r}
h(r)=-1.
\end{equation}
In the case of uncorrelated pancake arrangements in neighboring
layers $h(r)$ is the pair distribution function of the 2D liquid. Using
identity $\int d{\bf r}K({\bf r})=0$ we rewrite Eq.\ (\ref{zero}) as 
\begin{equation}
F_{v}(r)=-(n_{v}^{2}/2)\int d{\bf r}_{1}d{\bf R}h({\bf r}_{1})[\beta ({\bf r}
,{\bf R}-{\bf r}_{1}/2)-\beta ({\bf r},{\bf R}+{\bf r}_{1}/2)]^{2}
\end{equation}
Performing the integration over ${\bf R}$ and averaging over orientation of $
{\bf r}_{1}$ we finally obtain 
\begin{equation}
F_{v}(r)=-\pi n_{v}^{2}\int dr_{1}r_{1}h(r_{1})J(r,r_{1}),  \label{Fr_Kr}
\end{equation}
where the universal function $J(r,r_{1})$ is given by %\begin{equation}
%J(r,r_1)=(1/4\pi )\int d\Omega _{{\bf r}_1}\int d{\bf R}[\beta ({\bf 
%r},{\bf R}- {\bf r}_1/2)-\beta ({\bf r},{\bf R}+{\bf r}_1/2)]^{2}.
%\end{equation}
\begin{eqnarray}
&&J(r,r_{1})=2\pi r^{2}\left[ \ln (r_{1}/r)+2\right] ,\ \ \ r_{1}>r,
\label{Jr} \\
&&J(r,r_{1})=4\pi \left[ 2rr_{1}-r_{1}^{2}-(r_{1}^{2}/2)\ln (r/r_{1})\right]
,\ \ \ r_{1}<r.  \nonumber
\end{eqnarray}
The function $J(r,r_{1})$ and its first three derivatives are continuous
functions of $r$ and $r_{1}$ but the fourth derivative has jump $8\pi
/r_{1}^{2}$ at $r=r_{1}$. Eqs.\ (\ref{Fr_Kr}) and (\ref{Jr}) represent a
very important general relation connecting phase fluctuations with the
vortex density fluctuations in the Gaussian approximation for the pancake
liquid. The function $F_{v}(r)$ increases linearly at large $r$, 
\begin{equation}
F_{v}(r)\approx r/l_{0},\ \ \ a/l_{0}=-8\pi ^{2}\int_{0}^{\infty
}dxx^{2}h(x),\ \ \ r\gg a,  \label{Asimp1}
\end{equation}
indicating exponential decay of phase fluctuations at large distances. Here
we introduced the dimensionless distance $x=r/a$ with $a=n_{v}^{-1/2}$. For
small $r$ the function $F_{v}(r)$ increases approximately quadratically with 
$r$: 
\begin{equation}
F_{v}(r)\approx \pi (r/a)^{2}\left( \ln \frac{a}{r}+2+C_{2}\right) ,\ \ \
C_{2}=-2\pi \int_{0}^{\infty }dxxh(x)\ln x,\ \ \ r\ll a.
\end{equation}
Note that $h(x)$ depends on the temperature and pinning. The temperature
dependence is determined by the dimensionless parameter $T/E_{0}$ if the
effect of pinning is neglected. Respectively, the function $F_{v}(r/a)$ is
also temperature and pinning dependent.

\subsection{Effective Josephson coupling and plasma frequency}

The problem of calculating of ${\cal C}(B,T)$, which determines the
Josephson plasma frequency is now reduced to calculation of the integral
(\ref{Fr_Kr}) with pair distribution functions of the liquid, $h(r)$. 
These functions are not available analytically.  They can be calculated
from Monte Carlo simulations \cite{caillol} or (approximately) using the
density functional theory \cite{Menon}.  As the first approximation we
consider uncorrelated two dimensional liquids in layers.  For weakly
coupled superconductors the density correlations between the layers and
weak pinning can be taken into account perturbatively.  In the 2D
approximation the vortex system is equivalent to the one-component
two-dimensional Coulomb plasma, which was studied extensively in the
past \cite{caillol}.  The pair distribution function
$h_{2D}(r,B,T,E_{0})$ within this approximation has an exact scaling
property:
\begin{equation}
h_{2D}(r,B,T,E_{0})=h_{2D}(r/a,T/E_{0})
\end{equation}
i.e., it depends on magnetic field only through the spatial scale. As
follows from Eq.\ (\ref{Fr_Kr}) this scaling property is also transferred to
the function $F_{v}(r)$, $F_{v}(r)=F_{v}(r/a,T/E_{0})$. If, in addition,
fluctuations of the regular phase are neglected then at $B_{x}=0$ we obtain
from Eqs.$\;$(\ref{cosGvGs},\ref{GvFv}) the expression (\ref{cfun}) for $
{\cal C}$. Correlation properties of the liquid determine the field
independent universal function $f(T/E_{0})$ defined by Eq.~(\ref{fexpr}).
Therefore, the scaling property ${\cal C}\propto 1/B_{z}$ would be exact 
if we neglect (i) coupling between pancake liquids in different layers
(ii)pinning inside the layers, and (iii) regular phase fluctuations. 
All these effects are typically small in the pancake liquid phase.

Now, we further develop this approach and include fluctuations of the
regular phase, magnetic interlayer coupling and pinning.  First, we include the
effect of regular phase fluctuations. As the temperature approaches $T_{c}$
the role of these fluctuations of regular phase progressively increases. The
correlation function of regular phase $S_{r}({\bf r})$ is given by Eq.~(\ref
{scff}). We substitute (\ref{scff}) into Eq.\ (\ref{cosGvGs}) and take into
account that the Josephson coupling, $E_{J}\propto \lambda _{c}^{-2}$, and
the plasma frequency at zero field are renormalized by the phase
fluctuations. Their suppression is determined by the cosine factor at $B=0$, 
$C(T)$, which was estimated in Ref.~\onlinecite{glko} as 
\begin{equation}
C(T)=(\xi _{ab}/\lambda _{J})^{\alpha }.
\end{equation}
Using the scaling property of $S_{v}({\bf r})$ we obtain for magnetic field
along the {\it c} axis: 
\begin{equation}
\langle \cos [\varphi _{n,n+1}^{(v)}+\varphi _{n,n+1}^{(r)}]\rangle =f_{s}(T)
\frac{E_{0}(T)}{2T}\left( \frac{B_{J}}{B_{z}}\right) ^{1-\alpha },
\label{mr}
\end{equation}
where 
\[
f_{s}(T)=2\pi \int_{0}^{\infty }dxx^{1-2\alpha }S_{v}(x,T)
\]
again depends only on reduced temperature $T/E_{0}$ within the 2D
approximation.  Therefore, regular phase fluctuations reduce the power
index in the field dependence of ${\cal C}$ and make it temperature
dependent.  Numerical calculations, described below in Section
\ref{Sec-Numeric}, give $f_{s}(T)\approx 1.0-1.75 T/(2\pi E_{0})$ at
$\alpha = T/(2\pi E_{0})\lesssim 0.1$.  Using Eqs.~(\ref{cosGvGs}),
(\ref{scff}) and (\ref{zero}) we obtain the JPR frequency for
arbitrarily oriented magnetic field:
\begin{eqnarray}
&&\omega _{p}^{2}(B_{x},B_{z},T)=\omega _{0}^{2}(T)\frac{E_{0}}{2T}\left( 
\frac{B_{J}}{B_{z}}\right) ^{1-\alpha }f_{s}\left( \frac{2\pi sB_{x}}{\sqrt{
\Phi _{0}B_{z}}},T\right) ,  \label{result} \\
&&f_{s}(b,T)=2\pi \int_{0}^{\infty }dxx^{1-2\alpha }S_{v}(x,T){\mathrm J}_{0}(bx),
\label{int}
\end{eqnarray}
where ${\mathrm J}_{0}(x)$ is the Bessel function.

\subsection{Influence of interlayer coupling and pinning. Perturbative
estimate.}

We consider now the influence of weak interlayer magnetic coupling and weak
pinning. In both cases it is necessary to calculate a correction to the
density correlation function and substitute it into Eq.\ (\ref{Fr_Kr}). The
starting point of the calculation is the thermodynamic expression for the
density correlation function 
\begin{eqnarray}
K_{n-m}({\bf r}-{\bf r}^{\prime }) &\equiv &\left\langle \delta \rho _{n}(
{\bf r})\delta \rho _{m}({\bf r}^{\prime })\right\rangle   \label{KnmCont} \\
&=&\frac{1}{Z}\int D{\bf r}_{n\nu }\delta \rho _{n}({\bf r})\delta \rho _{m}(
{\bf r}^{\prime })\exp \left( -\frac{{\cal F}_{2D}+{\cal F}_{{\rm mc}}+{\cal 
F}_{{\rm pin}}}{T}\right).
\end{eqnarray}
Here we split the electromagnetic part of the energy (\ref{f2}) into the
intralayer two-dimensional part (${\cal F}_{2D}$) and the interlayer
magnetic coupling (${\cal F}_{{\rm mc}}$) part. It is convenient to
represent all energy contributions in continuous form: 
\begin{eqnarray}
{\cal F}_{2D} &=&\pi E_{0}\sum_{n}\int d{\bf r}d{\bf r}^{\prime }\ln \frac{a
}{\left| {\bf r}-{\bf r}^{\prime }\right| }\delta \rho _{n}({\bf r})\delta
\rho _{n}({\bf r}^{\prime }) \\
{\cal F}_{{\rm mc}} &=&\frac{1}{2}\sum_{n,m}\int d{\bf r}d{\bf r}^{\prime
}V_{{\rm mc}}({\bf r}-{\bf r}^{\prime },n-m)\delta \rho _{n}({\bf r})\delta
\rho _{m}({\bf r}^{\prime }) \\
{\cal F}_{{\rm pin}} &=&\sum_{n}\int d{\bf r}V_{{\rm pin}}({\bf r})\delta
\rho _{n}({\bf r})
\end{eqnarray}
with 
\begin{equation}
\left\langle V_{{\rm pin}}({\bf r})V_{{\rm pin}}(0)\right\rangle =W({\bf r}).
\end{equation}
Magnetic interaction between pancake vortices, $V_{{\rm mc}}({\bf r},n)$,
has been considered in Refs.~\onlinecite{BuzFei,ArtKrug,clem}. At small
distances $r,ns\ll \lambda _{ab}$ a simple analytical expression can be
obtained for $n>0$ 
\begin{equation}
V_{{\rm mc}}({\bf r},n)=\frac{\pi sE_{0}}{2\lambda _{ab}^{2}}\left[ \sqrt{
r^{2}+(ns)^{2}}-ns-ns\ln \frac{\sqrt{r^{2}+(ns)^{2}}+ns}{2ns}\right] .
\end{equation}
If both ${\cal F}_{{\rm mc}}$ and ${\cal F}_{{\rm pin}}$ are neglected
then $ K_{n-m}({\bf r})=K_{2D}({\bf r})\delta _{nm}$, with $K_{2D}({\bf
r})$ being the density correlation function of a 2D Coulomb plasma,
$K_{2D}({\bf r})=n_{v}\delta ({\bf r})+n_{v}^{2}h_{2D}({\bf r})$.

First, we obtain the correction to the density correlation function
induced by magnetic coupling, $\Delta _{{\rm mc}}K_{n-m}({\bf r}-{\bf
r}^{\prime })$.  Expanding ( \ref{KnmCont}) with respect to ${\cal
F}_{{\rm mc}}$ we obtain:
\begin{equation}
\Delta _{{\rm mc}}K_{n-m}({\bf r})=-\frac{1}{T} \int d{\bf r}_{1}d{\bf r}
_{2}V_{{\rm mc}}({\bf r}+{\bf r}_{1}-{\bf r} _{2},n-m)K_{2D}({\bf r}
_{1})K_{2D}({\bf r}_{2}).  \label{DeltaK_Magn}
\end{equation}
Interplane phase correlations are determined by the density
correlation function in neighboring layers, $|m-n|=1$ (see 
Eq.\ (\ref{DenCorr}) ). The integral (\ref
{DeltaK_Magn}) decreases rapidly at $r\gg a$, because the main contribution
comes from the distances $r_1,r_2$ of the order $a$ and $\int d{\bf r}
K_{2D}({\bf r})=0$. For $s\ll r <\lambda_{ab} $, the interaction potential
for pancakes in neighboring layers $V_{{\rm mc}}({\bf r},1)$ increases
linearly with $r$, $V_{{\rm mc}}({\bf r},1)\approx (\pi sE_{0}/2\lambda
_{ab}^{2})r$ and the correction to $K_{1}({\bf r})$ is: 
\begin{equation}
\Delta _{{\rm mc}}K_{1}({\bf r})=-\frac{\pi sE_{0}}{ 2\lambda _{ab}^{2}T}
\int d{\bf r}_{1}d{\bf r}_{2}\left(\left| {\bf r} +{\bf r}_{1}-{\bf r}
_{2}\right|-r\right) K_{2D}({\bf r}_{1})K_{2D}( {\bf r}_{2}).
\label{DeltaK_magn1}
\end{equation}
Using the scaling property of $K_{2D}({\bf r})$, we get a simple dimensional
estimate of this integral 
\begin{equation}
\Delta _{{\rm mc}}K_{1}({\bf r})/n_{v}^{2}=(saE_{0}/\lambda _{ab}^{2}T)F_{
{\rm mc}}(r/a),  \label{DeltaK_scal}
\end{equation}
where $F_{{\rm mc}}(x)$ is a dimensionless function of order unity. An exact
shape of this function can be obtained by numerical evaluation of the integral
in Eq.~(\ref{DeltaK_magn1}). Using the scaling estimate (\ref{DeltaK_scal})
we obtain the correction to ${\cal C}$ given by Eq.~(\ref{mr}) due to
magnetic coupling, $\Delta _{{\rm mc}}{\cal C}$: 
\[
\Delta _{{\rm mc}}{\cal C}=f_{{\rm mc}}\frac{s \Phi _{0}^{1/2}E_{0}^{2}B_{J} 
}{\lambda _{ab}^{2}T^{2}B^{3/2}} 
\]
with $f_{{\rm mc}}\sim 1.$ This correction is small because of the small
parameter $s/\lambda_{ab}\approx 0.01$ in atomically layered 
superconductors.

Let us turn now to the pinning-induced correction to the density
correlation function, $\Delta _{{\rm pin}}K_{n-m}({\bf r}-{\bf
r}^{\prime })=\Delta _{{\rm pin}}K_{0}({\bf r}-{\bf r}^{\prime
})\delta_{nm}$.  Expanding (\ref{KnmCont}) with respect to ${\cal
F}_{{\rm pin}}$ we obtain in the second order with respect to pinning
potential:
\begin{eqnarray}
&&\Delta _{{\rm pin}}K_{0}({\bf r}-{\bf r}^{\prime })=  \label{DeltaK_pin} \\
&&\frac{1}{2T^{2}}\int d {\bf r}_{1}d{\bf r}_{2}W\left( {\bf r}_{1}-{\bf r}
_{2}\right) \left[ \left\langle \delta \rho ({\bf r})\delta \rho ({\bf r}
^{\prime })\delta \rho ({\bf r}_{1})\delta \rho ({\bf r}_{2})\right\rangle
-K_{2D}({\bf r}-{\bf r} ^{\prime })K_{2D}({\bf r}_{1}-{\bf r}_{2})\right]. 
\nonumber
\end{eqnarray}
This correction is determined by the quartic correlation function of
density.  Such a perturbation approach is valid for weak pinning when
the pinning energy of a single vortex, $E_p\approx \sqrt{W(0)}$, is small in
comparison with the temperature.  
% Here the integral is calculated in the
% region of the vortex core with the area $\xi_{ab}^2$ around the point
% ${\bf r}$, and the maximum is found with respect to the coordinate of
% the vortex center ${\bf r} $.  
% The vortex-vortex interaction results in
% a strongly correlated vortex liquid when $T\ll E_0$, and, thus the
% effect of pinning on the vortex liquid with strong correlations may be
% treated as a perturbation at $E_p\ll T\ll E_0 $.  
An important observation following from Eq.~(\ref{DeltaK_pin}) is that,
in this temperature interval, the first pinning correction 
exactly coincides with the correction induced by the
extra interaction potential $W\left( {\bf r} _{1}-{\bf r} _{2}\right)
/T$.  The range of this potential coincides with the correlation length
of pinning, $r_{p}$, that is much smaller than the intervortex spacing. 
It means that this potential has almost no influence on the density
correlation function at $E_p\ll T$, because the contribution to the
correction comes only from the rare configurations when two particles
approach at distance of the order of $r_{p}$.  We can conclude,
therefore, that influence of pinning on the pair distribution function
of the liquid is negligible at $T\gg E_p$.  This conclusion is in
agreement with the more detailed study of influence of point pinning on
correlation properties of the pancake liquid using the replica approach
\cite{MenonReplica}.  Note, that strong pinning sites with $E_p\approx
E_0$, such as columnar defects produced by the heavy ion irradiation,
can not be treated perturbatively and may strongly influence density
correlations of the pancake liquid.

\subsection{Phase correlations in pancake liquid regime. Numerical
calculations}
\label{Sec-Numeric}

\begin{figure} \epsfxsize=3.4in \epsffile{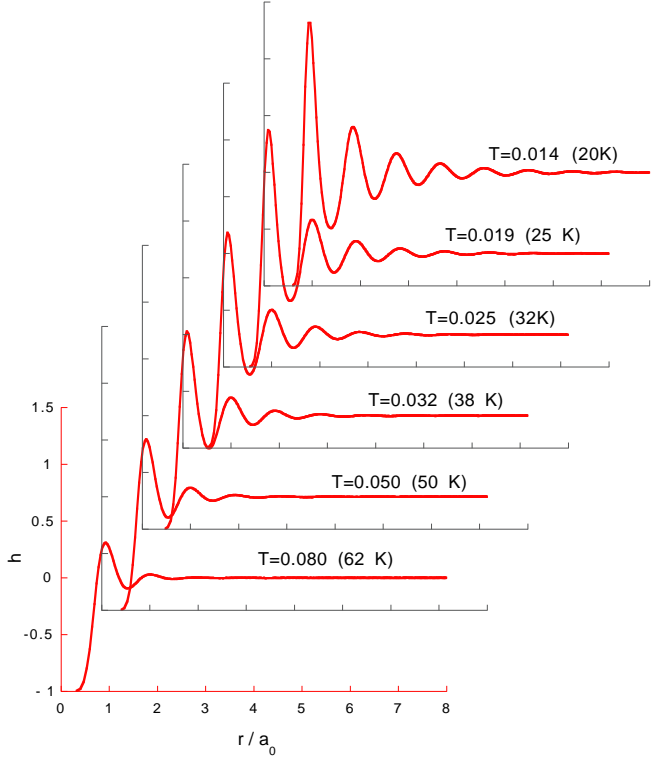}
\caption{Temperature evolution of pair distribution functions for a 2D
vortex liquid (Langevin dynamics simulations).  The plots are labeled
by reduced temperature measured in units $2\pi E_{0}$ (as in Refs.\
\protect\onlinecite{caillol}).  In parenthesis we also show
approximate values of corresponding real temperature for Bi-2212
assuming a temperature dependent London penetration depth
$\lambda_{ab}=200{\rm nm}/(1-(T/T_{c})^{2})^{1/2}$ with $T_{c}=90$K.}
\label{Fig-hr}
\end{figure}

\begin{figure}
\epsfxsize=3.4in \epsffile{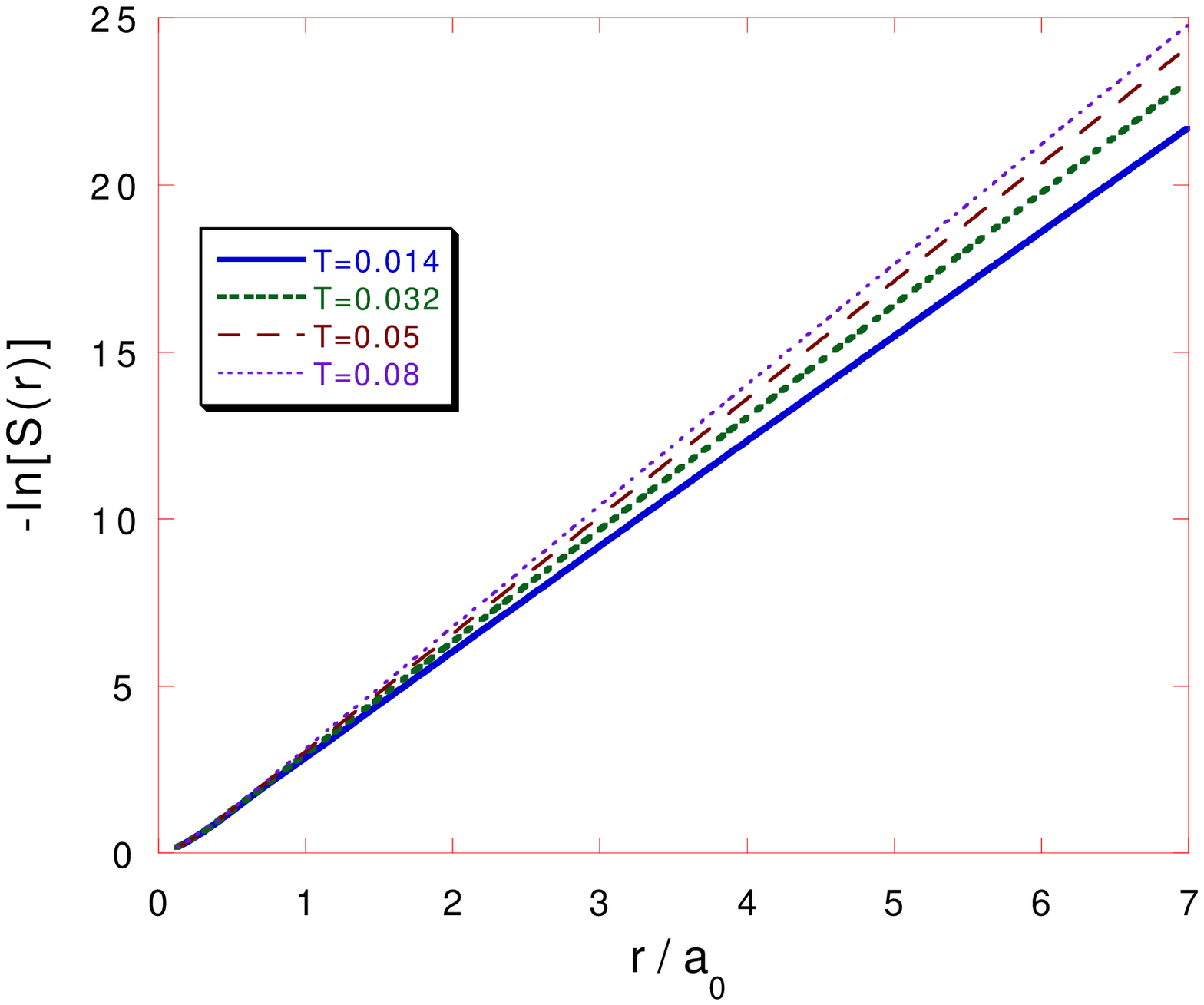} 
\caption{Phase correlation functions at different temperatures 
obtained by numerical integration of Eq.\ (\protect\ref{Fr_Kr}) with 
numerical pair distribution functions shown in Fig.\ \ref{Fig-hr}.}
\label{Fig-fr}
\end{figure}

\begin{figure}
\epsfxsize=3.4in \epsffile{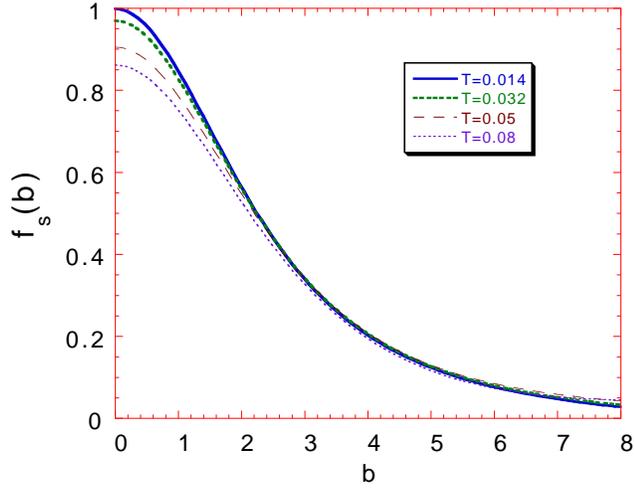} 
\caption{The Fourrier transform of
the phase correlation function $f_{s}(b)$ which determines the angular
dependence of the plasma frequency given by Eq.\ (\protect\ref{result}). 
The function is calculated by the Fourier-Bessel transformation
(\protect\ref{int}) using the phase correlation functions shown in Fig.\
\protect\ref{Fig-fr}.} 
\label{Fig-fb}
\end{figure}

\begin{figure}
\epsfxsize=3.4in \epsffile{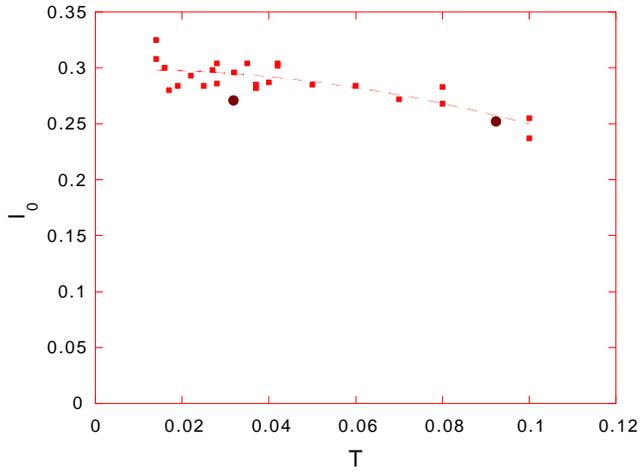} \caption{Temperature dependence of
the phase correlation length defined by Eq.\ (\ref{Asimp1}) in units
of lattice constant $a_{0}$.  Squares represent results of numerical
integration of Eq.~(\protect\ref{Asimp1}) using pair distribution
functions $h(r)$ generated from Langevin dynamics simulations.  Spread
of data is due to an error in the integration because of numerical noise in
$h(r)$ at large $r$.  Two filled circles represent correlation lengths
extracted directly from the phase correlation function, which was
obtained by Monte Carlo simulations of the frustrated XY model 
\protect\cite{Koshelev:PRB97}.}
\label{Fig-l0}
\end{figure}

In the decoupled liquid phase correlations can be calculated using the pair
distribution function $h(r)$ of the two-dimensional vortex system. The phase
correlation function $F_{v}({\bf r})$ is connected with $h(r)$ by relation (\ref
{Fr_Kr}). To study the behavior of phase correlations in the liquid state we
performed numerical simulations of the motion of $N_{v}=900$ point vortices
by numerical solution of the set of Langevin equations for the vortex
coordinates ${\bf R} _{i}$: 
\begin{equation}
{\frac{{{\rm d}{\bf R}_{i}}}{{{\rm d}t}}}=\sum_{i\ne j}{\bf f}_{v}({\bf R}
_{i}-{\bf R}_{j})+{\bf f}_{Li}(t).  \label{Langev}
\end{equation}
Here ${\bf f}_{v}({\bf r})$ is the intervortex interaction force, ${\bf
f} _{Li}(t)$ is the Langevin force, $\left\langle f_{Li\alpha
}(0)f_{Lj\beta }(t)\right\rangle =2T\delta _{\alpha \beta }\delta
_{ij}\delta (t)$.  The quantity $2\pi E_{0}=2s\Phi _{0}^{2}/(4\pi
\lambda_{ab} )^{2}$, and the lattice constant
$a_{0}=\sqrt{2/\sqrt{3}n_{v}} $ of the ideal lattice are taken as the
units of energy and length.  To facilitate numerical calculations we
model the real infinite-range $1/r$ interaction force by the
finite-range interaction force
$f_{v}(r)=(1/r)\left(1-{{r^{2}}/{r_{cut}^{2}}} \right) ^{2}$, cut off at
the finite length $r_{cut}$.  To minimize the effect of cutting we have
chosen the cutting length $r_{cut}=4.33$ to reproduce the shear modulus
$C_{66}=B\Phi_{0}/(8\pi \lambda_{ab})^{2}$ for the uncut interactions
(i.e., for $r_{cut}=\infty$).  We solve the Langevin equations
(\ref{Langev}) at different values of $T$ ranging from $0.006$ to $
0.1$.  Simulations give a melting transition at $T_{m}\approx 0.007$, in
agreement with the value of the melting point for the two-dimensional
Coulomb plasma \cite{caillol}.  A set of pair distribution functions,
$h(r)$, has been generated using averaging over a large number (
$\approx 2\cdot 10^{4}$ ) of vortex configurations (see Fig.\
\ref{Fig-hr}).  For each $h(r)$ we calculated the phase correlation
function, $S(r)$, integrating numerically Eq.\ (\ref{Fr_Kr}).  Several
phase correlation functions are presented in Fig.\ \ref{Fig-fr}.  Note
that the phase correlation function has surprisingly weak temperature
dependence.  To facilitate comparison with experiment, we also
calculated the function $f_{s}(b,T)$ that determines the angular
dependence of the plasma frequency via Eqs.\ (\ref{result}) and (\ref
{int}).  This function is plotted in Fig.\ \ref{Fig-fb}.  We also
calculated typical lengths, characterizing decay of $S(r)$: $l_{0}$
defined by Eq.\ (\ref{Asimp1}), $R$, and $R_{1}$, where the last two
lengths are defined by relations
\begin{equation}
R^{2} =\int_{0}^{\infty }rdrS(r),\; \; R_{1}^{2} =\int_{0}^{\infty
}r^{3}drS(r)/R^{2}.   \label{R2}
\end{equation}
Temperature dependence of the phase correlation length $l_{0}$ is shown in
Fig.~\ref{Fig-l0}. For comparison we also show two points extracted directly
from the phase correlation functions, which were generated using Monte Carlo
simulations of the uniformly frustrated XY model \cite{Koshelev:PRB97}. We
see that the results of the two models are in perfect agreement. We found
that, in spite of considerable change in shape of $h(r)$ with increasing
temperature, the phase correlation function does not change much. In
particular, we find that $l_{0}$ drops from $0.3\,a_{0}$ to $0.25\,a_{0}$
(see Fig.~\ref{Fig-l0}), $R^{2} \approx \left( 0.12-0.13\right) a_{0}^{2}$,
and $R_{1}^{2}\approx \left( 0.49-0.55\right) a_{0}^{2}$, ($R_{1}\approx 2R$
) in the investigated temperature range.

\subsection{Extracting the phase correlation and pair distribution functions
from angular dependence of the plasma frequency}

Measurements of dependence of the resonance frequency $\omega _{p}$ on $
B_{x} $ at fixed $B_{z}$ and $T$ allow one, in principle, to obtain the
phase correlation function $S(r)$ and the pair distribution function $h(x)$,
i.e., to extract quantitative information about correlations in the vortex
liquid. Using Eq.~(\ref{result}) the function $f(b)=f_{s}(b)/f_{s}(0)$ can
be found from the dependence $\omega _{p}(B_{x})$ as: 
\begin{equation}
f(b)=\frac{\omega _{p}^{2}(B_{x}=b\sqrt{\Phi _{0}B_{z}}/2\pi s)}{\omega
_{p}^{2}(B_{x}=0)}.  \label{fb}
\end{equation}
Usually the angular dependence of the resonance field $B_{r}(\theta)$ is
measured at fixed microwave frequency \cite{MatsPRB97,OngPhysC97} rather than $
\omega _{p}(B_{x})$. This dependence also can be used to extract $f(b)$: 
\begin{equation}
f\left[b=\frac{2\pi sB_{rx}(\theta)}{\sqrt{ \Phi _{0}B_{rz}(\theta)}}\right]=
\left[\frac{B_{rz}(\theta)}{B_{rz}(0)}\right] ^{1-\alpha}  \label{fb1}
\end{equation}

The function $F_{s}(x)$ may be found by the reverse Fourier transform from
Eq.~(\ref{int}), 
\begin{equation}
F_{s}(x,T)=-\ln \left[{\frac{f_s(0,T)x^{2\alpha }}{2\pi }} \int_{0}^{\infty
}dbf(b)b{\mathrm J}_{0}(bx) \right].
\end{equation}
In the final stage the density correlation function $h(x)$ according to Eq.~(
\ref{Fr_Kr}) is given by the relation: 
\begin{equation}
h(x)=-\frac{1}{8\pi x^{3}}\frac{d}{dx}\left\{ x^{3}\frac{d}{dx}\left[ x\frac{
d^{3}F_{s}(x)}{dx^{3}}\right] \right\} .  \label{de}
\end{equation}
Unfortunately it includes high order derivatives which makes the procedure
very difficult for practical realization.

It is more practical to compare the Fourier transform of the phase
correlation function $f(b)$, which can be directly extracted from the
angular dependence of resonance field via Eq.\ (\ref{fb1}), with the
theoretically predicted function obtained by numerical simulations (Fig.\ 
\ref{Fig-fb}). To make this comparison we used angular dependence of the
resonance field for an optimally doped Bi-2212 sample at 40 K from Fig.\ 3 of
Ref.\ \onlinecite{MatsPRB97}. We plot in Fig.\ \ref{Fig-fbcomp} the function $
B_{rz}(\theta)/B_{rz}(0)$ ($\alpha$ at this temperature can be neglected) vs 
$2\pi sB_{rx}(\theta)/\sqrt{\Phi _{0}B_{rz}(\theta)}$ and the theoretical $
f(b)$ at $T/2\pi E_{0}=0.032$, roughly corresponding to 40 K. One can see
that the experimental $f(b)$ is very close to the theoretical one. This
indicates that phase fluctuations in real Bi-2212 crystals are well described
by the pancake liquid model.

\subsection{Relation between the plasma frequency and the {\it c}-axis
conductivity in vortex liquid}

In this part we will establish a relation between $\omega_p$ and the {\it c}
-axis conductivity $\sigma_c$ in the vortex liquid which allows us to
estimate the characteristic frequency of phase slips, $\omega_{ps}$, from
experimental data for $\omega_p$ and $\sigma_c$. For the plasma frequency we
obtain: 
\begin{eqnarray}
&&\omega_p^2(B)=(4\pi sJ_0^2/\epsilon_cT)\int d{\bf r}S({\bf r},B,t=0), \\
&&S({\bf r},B,t)=\langle\cos[\varphi_{n,n+1}({\bf r},t)-\varphi_{n,n+1}(0,0) 
]\rangle,  \label{sig}
\end{eqnarray}
The {\it c}-axis conductivity below $T_c$ consists of the quasiparticle
and Josephson contributions.  The conductivity due to the interlayer
Josephson currents is finite in the liquid phase because of the phase
slips induced by pancake thermal motion \cite{KoshPRL96A}.  When the
temperature is decreased at fixed field, the system smoothly crosses
over from the region of dominating quasiparticle transport to the region
of dominating Josephson transport.  We consider the latter region, where
the conductivity is related to the correlation function of the Josephson
currents via the Kubo formula
\begin{equation}
\sigma_c=(sJ_0^2/T)\int_0^{\infty}dt\int d{\bf r}\langle\sin
\varphi_{n,n+1}(0,0)\sin\varphi_{n,n+1}({\bf r},t)\rangle= (sJ_0^2/2T)\int d 
{\bf r}dtS({\bf r},t).  \label{kubo}
\end{equation}
Now the characteristic frequency $\omega_{ps}$ involved in the {\it c} axis
conductivity may be estimated using the relation 
\begin{equation}
\omega_{ps}=\frac{\int d{\bf r}S({\bf r},0)}
{\int_0^{\infty} dt\int d{\bf r}S({\bf r },t)}
=\frac{\epsilon_c\omega_p^2}{8\pi\sigma_c},
\end{equation}
if both conductivity and plasma frequency are known at a given temperature and
magnetic field. 

In Bi-2212 crystals with $T_c\approx 85$ K a $c$-axis conductivity
$\sigma_{c}\approx 0.5$ ohm$^{-1}$cm$^{-1}$ was found in Ref.\
\onlinecite{moroz} at $T=67$ K and $B_z=0.3$ T. JPR frequency in a
crystal with similar $T_c$ is $\omega_p/2\pi\approx$ 120 GHz at 4.2 K
and $\approx 90 $ GHz at 67 K \cite{gai}.  From these data, using
Eq.~(\ref{result}) we estimate $\omega_p/2\pi\approx 10$ GHz at $T=67$ K
and $B_z=0.3$ T. With $ \epsilon_c\approx 12$ we obtain
$\omega_{ps}/2\pi\approx 0.2$ GHz at $B_z=0.3 $ T.
% , and frequency 
% of phase slips only slightly increases with $B_z$ (both $\omega_p^2$ and 
% $\sigma_c$ decrease approximately as $1/B_z$ up to the field $B_z=1.5$ T).  
Thus we estimate $\omega_{ps}/\omega_p\approx 0.02$ which justifies the
static approach for calculation of plasma frequency.

Consider now influence of the in-plane field on the c-axis conductivity.  In the
lowest order in Josephson coupling, we split $[\varphi_{n,n+1}(0,0)-
\varphi_{n,n+1}({\bf r},t)]$ into the contribution induced by pancakes
and that caused by the unscreened parallel component $B_{x}$. 
Assuming that $B_{x}$ is along the $x$ axis, we obtain a simple
expression for the contribution of the parallel component to the phase
difference:
\begin{eqnarray}
&&\varphi_{n,n+1}(0,0)-\varphi_{n,n+1}({\bf r},t)\approx  \nonumber \\
&&[\varphi_{n,n+1}(0,0)-\varphi_{n,n+1}({\bf r},t)]_{B_{x}=0}- 2\pi
sB_{x}y/\Phi_0.  \label{me}
\end{eqnarray}
Inserting this expression into Eqs.~(\ref{sig}) and (\ref{kubo}) we obtain 
\begin{equation}
\sigma_c(B_{z},B_{x})=(\pi s J_0^2/T)\int drr\tilde{G}
(r,B_{z}) {\mathrm J}_0(2\pi s B_{x}r/\Phi_0),  \label{me1}
\end{equation}
where 
\begin{equation}
\tilde{G}({\bf r},B_{z})=\int_0^{\infty} dtS({\bf r},t,B_{z}).
\end{equation}
The function $S({\bf r},t)$ describes the dynamics of the phase difference
caused by mobile pancakes. If $g(B_{z},B_{x})=\sigma_c(B_{
\bot},B_{x})/\sigma_c(B_{z},0)$ is known in the vortex liquid,
the correlation function $G(r)=\tilde{G}(r)J_0^2\Phi_0^2/4\pi
sT\sigma_c(B_{z},0)$ may be found using the inverse Fourier-Bessel
transform: 
\begin{equation}
G(r,B_{z})=\int dB_{x} B_{x}g(B_{x},B_{z}){\mathrm J}_0(2\pi s B_{x}r/\Phi_0).
\label{tr}
\end{equation}
This procedure was realized in Ref.~\onlinecite{moroz} and the function 
$G(r,B_z)$ was found for pristine and irradiated Bi-2212 crystals.

\section{Line shape of plasma resonance}

\subsection{Formulation of the problem and general relations}

To find the line shape we need an equation for the phase difference
which describes time variations of the interlayer phase difference and
intralayer charges in the presence of an alternating electric field
applied along the {\it c} axis.  We will use the simplest version of
this equation (\ref {LinPhasDyn}), which is similar to the
Schr\"{o}dinger equation.  Thus the problem of inhomogeneous line
broadening is very similar to the problem of the density of states of a
quantum particle in a random potential, and one can use the same
techniques.  The role of the random potential in our problem is played
by $u_{n}({\bf r})=\cos [\varphi _{n,n+1}^{(0)}({\bf r})]-{\cal C}$. 
Neglecting the Josephson coupling, we obtain:
\begin{equation}
\langle u_{n}({\bf r})u_{m}(0)\rangle =\frac{S({\bf r})}{2}\delta _{nm}.
\end{equation}
where $S({\bf r})$ is the phase correlation function (\ref{coscorr}).
At large frequencies the random potential can be treated as a small 
perturbation. The exact criterion will be formulated below.   
Perturbative analysis of Eq.~(\ref{LinPhasDyn}) can be conveniently
performed using the Green function formalism. The Green function $\tilde{G}
_{mn}\left( {\bf r},{\bf r}^{\prime };E\right) $ is defined as the solution
of the equation: 
\begin{equation}
\sum_{m}\left\{ [E-{\cal C}+u_{n}({\bf r})]\delta _{nm}+\lambda
_{J}^{2}L_{nm}\nabla ^{2}\right\} \tilde{G}_{mn^{\prime }}\left( {\bf r}, 
{\bf r}^{\prime };E\right) =\delta _{nn^{\prime }}\delta \left( {\bf r}-{\bf 
r}^{\prime }\right) .
\end{equation}
with $E=\omega ^{2}/\omega _{0}^{2}$. The homogeneous part of the
alternating averaged phase is connected with $\tilde{G}_{nn^{\prime }}\left( 
{\bf r}, {\bf r}^{\prime };E\right) $ by the relation: 
\begin{equation}
\langle \varphi _{n,n+1}^{\prime }(\omega )\rangle =-\frac{i\omega {\cal
D}_{z}(\omega )}{4\pi J_{0}}\sum_{n^{\prime }}\int d{\bf r}^{\prime
}\left \langle \tilde{G}_{nn^{\prime }}\left( {\bf r},{\bf r}^{\prime };\frac{
\omega ^{2}}{\omega _{0}^{2}}\right ) \right \rangle =-\frac{i\omega {\cal
D}_{z}(\omega )}{4\pi J_{0}}G\left( {\bf k}=0,q=0;\frac{ \omega
^{2}}{\omega _{0}^{2}} \right ) ,
\end{equation}
where $G({\bf k},q;E)$ is the Fourier component of $G_{nn^{\prime }}\left( 
{\bf r}-{\bf r}^{\prime };E\right) =\langle \tilde{G}_{nn^{\prime }}\left( 
{\bf r},{\bf r}^{\prime };E\right) \rangle $. The homogeneous part of the
averaged electric field between layers $n$ and $n+1$ is: 
\begin{equation}
\langle E_{z}(\omega )\rangle =\frac{i\omega ^{2}\Phi _{0}}{2\pi
cs}\langle \varphi _{n,n+1}^{\prime }(\omega )\rangle =\frac{1}{\epsilon
_{c}}\frac{ \omega ^{2}}{\omega _{0}^{2}}G\left( 0,0;E=\frac{\omega
^{2}}{\omega _{0}^{2}}\right) {\cal D}_{z}(\omega ).
\end{equation}
Therefore, the dielectric function which describes the response to a
homogeneous external electric field is connected with the Green function as: 
\begin{equation}
\frac{1}{\epsilon (\omega )}=\frac{1}{\epsilon _{c}}\frac{\omega ^{2}}{
\omega _{0}^{2}}G\left( 0,0;E=\frac{\omega ^{2}}{\omega _{0}^{2}}\right) .
\end{equation}
The JPR absorption and shape of the JPR line is determined by ${\rm Im}
[1/\epsilon (\omega )]$.

\subsection{Diagrammatic expansion with respect to the random potential $
u_{n}({\bf r})$}

The Green function at $u({\bf r})=0$ is 
\begin{equation}
G_{0}({\bf k},q;E)=\frac{1}{E-\kappa ^{2}({\bf k},q)},\ \ \ \ \kappa ^{2}( 
{\bf k},q)={\cal C}+\frac{\lambda _{J}^{2}k^{2}}{2(1-\cos q)+s^{2}/\lambda
_{ab}^{2}}.
\end{equation}
This function determines the spectrum of the phase collective mode $\omega ( 
{\bf k},q)=\omega _{0}\kappa ({\bf k},q)$. Using a standard diagrammatic
approach, we can write: 
\begin{equation}
G({\bf k},q;E)=[E-\kappa ^{2}({\bf k},q)-\Sigma ({\bf k},q;E)]^{-1},
\label{green}
\end{equation}
where $\Sigma ({\bf k},q;E)$ is the self-energy. In the lowest order with
respect to $u({\bf r})$ (the Born approximation for scattering) we obtain: 
\begin{equation}
\Sigma \left( {\bf k},q;E\right) =\frac{1}{2}\int \frac{d{\bf k}^{\prime
}dq^{\prime }}{(2\pi )^3}G_{0}\left( {\bf k}^{\prime },q^{\prime };E\right)
S({\bf k}- {\bf k}^{\prime }).  \label{SelfEnBorn}
\end{equation}
The imaginary part of the self-energy, $\Sigma _{2}({\bf k},q;E)$,
determines line broadening, while the real part, $\Sigma _{1}({\bf k},q;E)$,
determines the shift of the resonance center due to inhomogeneities. For $E- 
{\cal C}\ll \lambda _{J}^{2}/a^{2}$, one can neglect the ${\bf k}$
-dependence of $S({\bf k})$ and we obtain: 
\begin{eqnarray}
\Sigma _{2}({\bf k},q;E) &=&(S_{0}/16\pi ^{2})\int d{\bf k}dq\delta \left[
E- {\cal C}-\kappa ^{2}({\bf k},q)\right] =(S_{0}/4\lambda _{J}^{2})\Theta
(E- {\cal C}),  \label{imsig} \\
\Sigma _{1}({\bf k},q;E) &=&-\frac{S_{0}}{4\pi \lambda _{J}^{2}}\ln \frac{
\lambda _{J}^{2}/a^{2}}{\left| E-{\cal C}\right| },
\end{eqnarray}
where $\Theta (x)=1$ at $x>0$, and vanishes otherwise. $S_{0}$ is the
Fourier component of the function $S({\bf r})$ at ${\bf k}=0$, i.e. 
\begin{equation}
S_{0}=\int d{\bf r}\left\langle \cos \left[ \varphi _{n,n+1}^{(0)}({\bf r}
)-\varphi _{n,n+1}^{(0)}(0)\right] \right\rangle .
\end{equation}
It is of the order $a^{2}$ in both the liquid phase and the vortex glass
phase with strong disorder.  In addition, in the liquid phase, the high
temperature expansion gives the relation $S_{0}=2{\cal C}T/E_{J}$, which
allows us to connect the width of the line with its center.  Parameter
$\Sigma _{\infty}\equiv S_{0}/4\lambda _{J}^{2} = {\cal C}T/(2E_{0})$
gives the reduced scattering rate due to the random potential. 
Perturbative expansion with respect to the random potential works if
$E- {\cal C}\gg \Sigma _{\infty}$.  In the real units this criterion
is given by $(\omega^{2}- \omega_{p}^{2})/\omega_{p}^{2}\gg T/2E_{0}$.
In spite of strong local variations of the random potential $u({\bf r})$,
which are much larger than ${\cal C}$, the parameter $\Sigma_{\infty}$
is much smaller than ${\cal C}$.  The reason is that typically the
{\it ac} phase $\phi_{n,n+1}^{\prime}({\bf r})$ varies smoothly in space
and effectively averages out these short-range variations.

The dielectric function can be now represented as
\begin{equation}
\frac{1}{\epsilon (\omega )}=\frac{1}{\epsilon _{c}}\frac{\omega ^{2}}{
\omega ^{2}-\omega _{p}^{2}\left[ 1-s_{1}(\omega )\right] -i\omega
_{b}^{2}\Theta (\omega ^{2}-\omega _{p}^{2})}.
\end{equation}
Here $\omega _{b}=(T/2E_{0})^{1/2}\omega _{p}$ is the line width for the
vortex liquid at the right-hand side of the resonance line, and $
s_{1}(\omega )=(T/2\pi E_{0})\ln [c_{0}^{2}/a^{2}(\omega ^{2}-\omega
_{p}^{2})]$.  The shape of the resonance line in the Born approximation
is given by the expression:
\begin{equation}
{\rm Im}\left[ \frac{1}{\epsilon (\omega )}\right] =\frac{1}{\epsilon _{c}} 
\frac{\omega ^{2}\omega _{b}^{2}}{\left ( \omega ^{2}-\omega _{p}^{2}\left[
1-s_{1}(\omega )\right] \right ) ^{2}+\omega _{b}^{4}}\Theta (\omega
^{2}-\omega _{p}^{2}),
\end{equation}
The Born approximation gives only the right-hand side of the resonance
line because it describes transformation of the homogeneous plasma mode
(with $k=0 $ and $q=0$) into the inhomogeneous phase collective modes
with nonzero momenta $k\lesssim a/\lambda _{J}^{2}$ and $q$ of the order
unity in the presence of the random coupling $\cos \varphi
_{n,n+1}^{(0)}({\bf r})$.  Estimate for values $k$ involved in the line
broadening follows from the relation $\kappa ^{2}( {\bf k},q)\approx
S_{0}/4\lambda _{J}^{2}$ at values $ q$ of the order unity.  This means
that the main part of the JPR line near its center at $\omega>\omega_p$
comes from averaging over distances $ R\lesssim \lambda _{J}^{2}/a$ in
the vortex liquid state.  Fluctuations of potential at smaller distances
result in a far right-hand wing of the JPR line.  Note that $\omega
_{b}$ in the liquid phase is small compared with the frequency of the
line center due to the small factor $\mu=T/2E_{0}$.

The absorption as function magnetic field is given by 
\begin{equation}
{\rm Im}\left[ \frac{1}{\epsilon (\omega ,B)}\right] =\frac{1}{\epsilon _{c}}
\frac{\mu B_{r}B}{\left[ B-B_{r}(1-s_{1})\right] ^{2}+\left( \mu
B_{r}\right) ^{2}}\Theta \left( B-B_{r}\right) ,  \label{fi}
\end{equation}
where $B_{r}=f\omega _{0}^{2}E_{J}\Phi _{0}/2T\omega ^{2}$ is the
resonance magnetic field in the liquid phase.  Eq.~(\ref{fi}) is
valid at $B-B_{r}\gg \mu B_{r}$, and, thus the term $(\mu B_{r})^{2}$ in
the denominator may be dropped.  Then the dependence of absorption on
$B$ is proportional to $ B/(B-B_{r})^{2}$, and this dependence is in
agreement with experiment \cite{MatsPRL95}.

\subsection{Self-consistent Born Approximation}

\begin{figure}
\epsfxsize=3.2in \epsffile{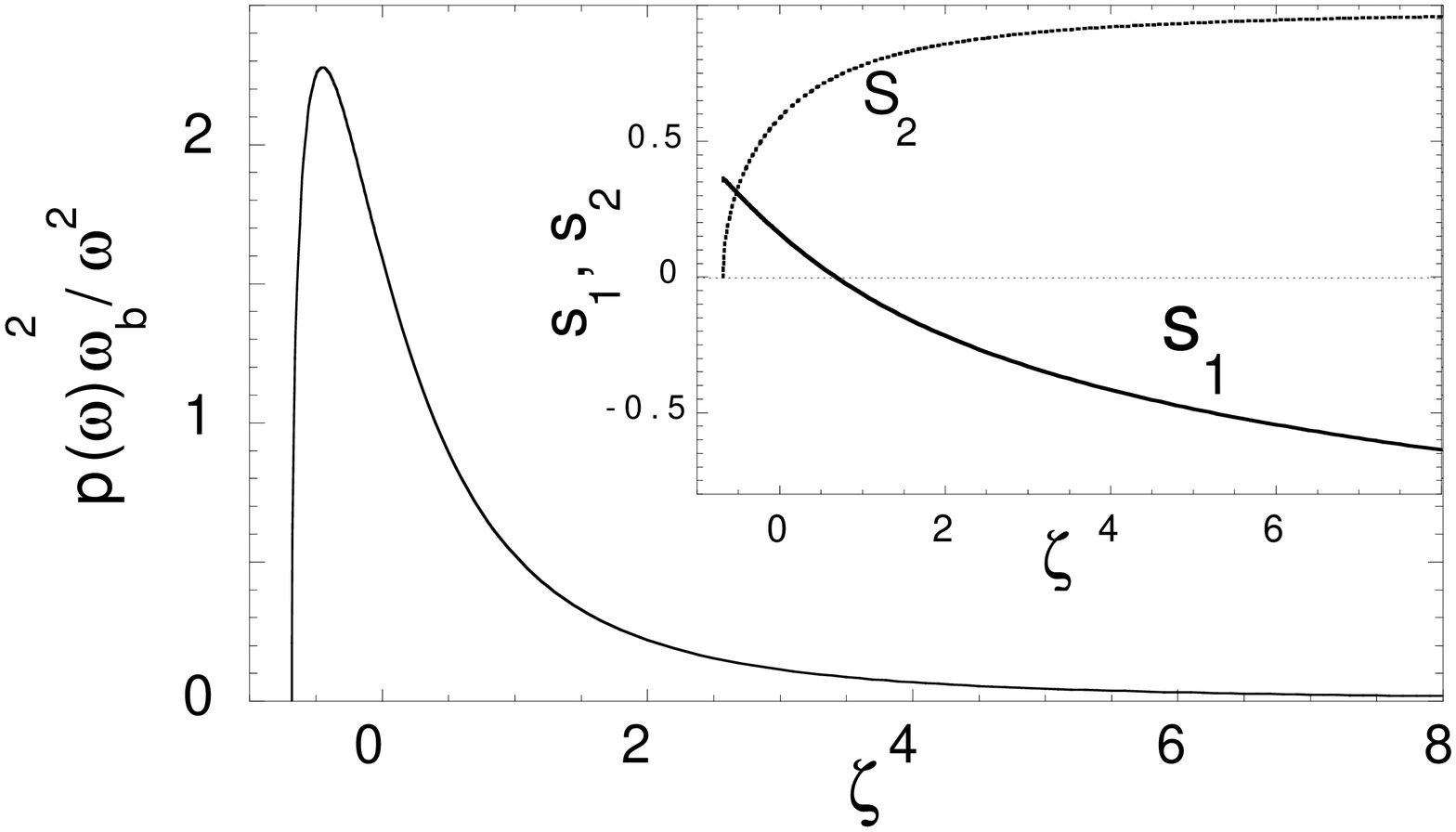} \caption{Plot of reduced JPR
lineshape $p(\omega)\omega_b^2/\omega^2$ vs reduced frequency $\zeta$
obtained within SCBA (Eqs.\ (\protect \ref{diel_freq}), (\protect
\ref{SCBA2}), and (\protect \ref{SCBA1})).  The insert shows frequency
dependencies of the dimensionless self energies $s_1$ and $s_2$.}
\label{Fig-SCBorn}
\end{figure}

Perturbative analysis of the previous Section can be improved by using
the renormalized Green function ~(\ref{green}) in the Born expression
for the self energy (\ref{SelfEnBorn}).  This approximation, known as the
self-consistent Born approximation, leads to the following equation for
the self-energy
\begin{equation}
\Sigma \left( {\bf k},q;E\right) =\frac{S_{0}}{2(2\pi )^{3}}\int d{\bf k}
^{\prime }dq^{\prime }\frac{1}{E-\kappa ^{2}({\bf k}^{\prime },q^{\prime
})-\Sigma \left( {\bf k}^{\prime },q^{\prime };E\right) }.  \label{SelfCons}
\end{equation}
Introducing dimensionless function $s_{1}(\zeta )$ and $s_{2}(\zeta )$ of
the dimensionless variable $\zeta =(E-{\cal C})/\Sigma _{\infty }+(2/\pi
)\ln (\lambda _{J}^{2}/a^{2})$ with $\Sigma _{\infty }\equiv
S_{0}/4\lambda_{J}^{2}$ as: 
\begin{equation}
\frac{\Sigma _{2}}{\Sigma _{\infty }}=s_{2}(\zeta ),\ \ \frac{\Sigma _{1}}{
\Sigma _{\infty }}=-\frac{2}{\pi }\ln \frac{\lambda _{J}^{2}}{a^{2}}
-s_{1}(\zeta )  \label{SelfEnSc}
\end{equation}
we obtain the following equations for these functions 
\begin{eqnarray}
s_{2} &=&\frac{1}{2}+\frac{1}{\pi }\arctan \frac{\zeta +s_{1}}{s_{2}} 
\label{SCBA2} \\
s_{1} &=&-\frac{1}{2\pi }\ln \left[ (\zeta +s_{1})^{2}+s_{2}^{2}\right]
\label{SCBA1}
\end{eqnarray}
The resonance absorption $p(\omega )$ can be written in terms of these
functions as: 
\begin{eqnarray}
&&p(\omega )\omega _{b}^{2}/\omega ^{2}=f\left[ (\omega ^{2}-\tilde{\omega}
_{p}^{2})/\omega _{b}^{2}\right] ,  \label{diel_freq} \\
&&f(\zeta )=\frac{s_{2}(\zeta )}{[\zeta +s_{1}(\zeta 
)]^{2}+s_{2}^{2}(\zeta )}. 
\end{eqnarray}
Here $\tilde{\omega}_p$ is the resonance frequency shifted by the random
Josephson coupling, $\tilde{\omega}_p^2=\omega_p^2(1-(T/\pi
E_0)\ln(B/B_{J})$).  This negative shift is due to the second order
perturbative correction to the ground state ``energy'', similar to the
well known result of quantum mechanical perturbation theory.  The
scaling representation (\ref{diel_freq}) is exact, i.e., the shape of
the line is fully determined by two dimensionless functions
$s_{1}(\zeta)$ and $s_{2}(\zeta)$.  Eqs.\ (\ref{SCBA2}) and
(\ref{SCBA1}) allow us to calculate these functions
approximately.  Fig.\ \ref{Fig-SCBorn} shows dependencies of the reduced
JPR absorption $p(\omega )\omega _{b}^{2}/\omega ^{2}$ and the functions
$s_{1,2}$ on the reduced frequency $\zeta $.  The self-consistent Born
approximation corresponds to summation of the perturbation series, which
includes only certain class of diagrams.  Omited terms have the same
order and this approximation does not describe quantitatively the line
shape in the region of maximum and below.  In particular, within this
approximation $\Sigma _{2}$ and resonant absorption still vanish at
$E<{\cal C}-\Sigma _{\infty }/\pi $.  In the real situation, there is a
long absorption tail due to the localized plasma modes arising from
exponentially rare fluctuation suppression of coupling at large areas. 
This tail is very similar to the Lifshitz tail in the density of states
in inhomogeneous semiconductors \cite{LifTail,Lifshitz}.

\subsection{Lifshitz Tail}

To calculate the exponential tail due to localized plasma modes (left-hand
side of the resonance line) one has to use the method of optimal 
fluctuation\cite{LifTail}.  We
will follow the derivations present in the book of Lifshitz, Pastur, and
Gredeskul \cite{Lifshitz}. From now on we make the replacement $E\rightarrow
E-{\cal C}$. The resonant absorption is determined by the averaged spectral
density 
\begin{equation}
\rho (E)=\frac{1}{\pi }\int d{\bf r}^{\prime }\sum_{n^{\prime }}{\rm Im}
\left[ G_{nn^{\prime }}\left( {\bf r}-{\bf r}^{\prime };E-i\delta 
\right) \right] ,
\end{equation}
which can be represented as: 
\begin{equation}
\rho (E)=\int d{\bf r}\sum_{n}\int Du({\bf r})P\{u_{n}({\bf r})\}\sum_{\nu
}\Psi _{\nu }(0,0;u)\Psi _{\nu }({\bf r},n;u)\delta \left( E-{\cal E}_{\nu
}\{u\}\right) ,
\end{equation}
where $\Psi _{\nu }({\bf r},n;u)$ and ${\cal E}_{\nu }$ are eigenfunction
and eigenvalue of the state $\nu $ in the system with the potential $u_{n}( 
{\bf r})$. The probability of realization of this potential is $P\{u_{n}( 
{\bf r})\}=\exp [-{\cal H}\{u({\bf r})\}]$. The scale of variations of typical
eigenfunctions $\Psi _{\nu }({\bf r},n;u)$ for not too small ${\cal E}_{\nu }
$ is much larger than the scale of variation of $u_{n}({\bf r})$. 
This allows us
to treat $u_{n}( {\bf r})$ as a short ranged Gaussian random variable and
approximate ${\cal H}\{u_{n}({\bf r})\}$ as 
\begin{equation}
{\cal H}\{u_{n}({\bf r})\}=\sum_{n}\int d{\bf r}\frac{\left[ u_{n}({\bf r}) 
\right] ^{2}}{S_{0}}.
\end{equation}
This is a good approximation to find the line shape not far away from the
line center because the eigenvalue ${\cal E}_{0}$ not far away from ${\cal C}
$ comes from the effective random potential $u_{n}({\bf r})$ averaged over
large length scales $R$. This approximation breaks down only for very large
and highly improbable fluctuations of the random potential. With exponential
accuracy in the tail region we have $\rho (E)\propto \exp (-\Phi (E))$ with 
\begin{equation}
\Phi (E)=\min_{u}{\cal H}\{u_{n}({\bf r})\}|_{{\cal E}_{0}\{u\}=E},
\label{PhiE}
\end{equation}
where ${\cal E}_{0}\{u\}$ is the lowest eigenvalue at given
$u_{n}({\bf r})$.  Following the standard line of reasoning
\cite{Lifshitz}, which we outline in the Appendix, we obtain that
$\Phi (E)$ is determined by solution of the nonlinear dimensionless
eigenvalue problem:
\begin{equation}
-\nabla ^{2}\psi _{n}+\nabla _{n}^{2}\psi _{n}^{3}=-\epsilon _{0}\nabla
_{n}^{2}\psi _{n}({\bf r}).  \label{DimEq}
\end{equation}
with condition $\psi_{n}(0)=1$. $\Phi (E)$ can be found from solution of
this equation as: 
\begin{equation}
\Phi (E)=\frac{\lambda _{J}^{2}E}{\epsilon _{0}S_{0}}\psi _{4}.
\label{PhiE1}
\end{equation}
with $\psi _{4}=\sum_{n}\int d{\bf r}\psi _{n}^{4}(r)$.
Numerical solution of Eq.~(\ref{DimEq}) gives $\epsilon _{0}=-0.1642$ and $
\psi _{4}=2.4178$. Asymptotic solution at large distances corresponds to a
dipole field: 
\begin{equation}
\psi _{n}({\bf r})=0.224\frac{\left| \epsilon _{0}\right| r^{2}-2n^{2}}{
\left( \left| \epsilon _{0}\right| r^{2}+n^{2}\right) ^{5/2}}
\end{equation}
because the time derivative of the phase difference $\varphi _{n,n+1}({\bf r}
)=\Psi _{n}({\bf r})$ gives the electric field $E_{z}$ between layers $n$
and $n+1$. This electric field of the dipole type describes plasma
oscillations with low frequencies when opposite charges are distributed
between layers in the regions where Josephson coupling is below the average
value due to fluctuations in the vortex density.

From Eq.~(\ref{PhiE1}) we obtain $\Phi (E)=14.72\lambda _{J}^{2}E/S_{0}$,
which corresponds to the low-frequency tail (low-magnetic-field tail) of the
dielectric function: 
\begin{equation}
{\rm Im}\left[ \frac{1}{\epsilon(\omega )}\right] \propto \exp \left[ -7.36 
\frac{E_0(\omega_p^2-\omega^2)}{T\omega_p^2}\right]
\end{equation}
at fixed magnetic field ($\omega_p$ depends on $B$). At fixed frequency, the
imaginary part of the inverse dielectric function as a function of magnetic
field is: 
\begin{equation}
{\rm Im}\left[ \frac{1}{\epsilon(B )}\right] \propto\exp\left(-3.68\frac{
B_r-B}{\mu B_r}\right).
\label{TailB}
\end{equation}
where $\mu = T/2E_{0}$ and the frequency dependent resonance field
$B_{r}$ is defined after Eq.\ (\ref{fi}).  In real units the
coarse-grained optimum cosine fluctuation at $E={\cal C} -\omega
^{2}/\omega _{0}^{2}\ll {\cal C}$ is:
\begin{equation}
u_{n}({\bf r})=\cos \varphi _{n,n+1}-{\cal C}=\frac{E}{\left| \varepsilon
_{0}\right| }\psi _{n}^{2}\left(\frac{{\bf r}}{r_{E}}\right), \ \ \
r_{E}=\lambda _{J}\sqrt{\frac{\left| \epsilon_{0}\right| }{E}}.  \label{fi1}
\end{equation}
Thus both sides of the resonance line are characterized by the width $
\omega_b\approx(T/E_0)^{1/2}\omega_p$ in the vortex liquid phase at
$B_z\gg B_J$.  Comparing the shape of the left-hand side of the
resonance line, Eq.~(\ref{TailB}), with that of the right-hand side,
Eq.~(\ref{fi}), we see that the low-field side is sharper in agreement 
with experiment \cite{MatsPRL95,MatsPRL97,KadowPRB97}.  

By using Eq.~(\ref{fi}) and data of Matsuda {\it et al.} \cite{MatsPRL95} for the
resonance line in the liquid phase at $T=36$ K, we may estimate the
parameter $\mu\approx 0.06$, while our theoretical estimate $\mu=T/2E_0(T)$
gives $\approx 0.13$ at $\lambda_{ab}=2500$ \AA\ \ and $s=15.6$ \AA. We note
that, actually, Eq.~(\ref{fi}) may be invalid near $B=B_r$, and, thus, our
estimate of $\mu$ from experimental data is correct by order of magnitude
only. The line was found to be anisotropic with a sharper left-hand side in
agreement with Eqs.~(\ref{fi}) and (\ref{fi1}). Thus, our model gives at
least qualitative explanation of the JPR line form.

\section{Conclusions}

We have obtained a quite complete description of the JPR for the vortex
liquid phase in magnetic fields $B_{z}\gg B_{J}$. We related the JPR
frequency in the presence of a tilted magnetic field, $\omega
_{p}(B_{x},B_{z})$, to the correlation function of the pancake liquid.
Numerical results for the density correlation function of the
two-dimensional Coulomb plasma allow one to predict the field dependence of
the plasma frequency or {\it vice versa}; experimental data for $\omega
_{p}(B_{x},B_{z})$ allow one to obtain information on the density
correlations in the pancake liquid.

We have explained the line shape of the JPR in the liquid phase by
inhomogeneous broadening caused by disorder in pancake positions along
the {\it c} axis.  It is the excitation of the phase collective modes
with nonzero momenta by a homogeneous-in-space alternating electric
field in the presence of an inhomogeneous vortex distribution that leads
to line broadening of the JPR line at $\omega>\omega_p$.  Rare
fluctuation suppressions of coupling in large areas due pancake density
fluctuations result in the left-hand side ($ \omega<\omega_p$) of the
resonance line broadening.  These mechanisms lead to a JPR line that
agrees well with experimentally observed asymmetric line shape.  Due to
this inhomogeneous broadening of JPR line mechanisms associated with
dissipation due to quasiparticles remain hidden when measurements are
made in a disordered vortex state.
% Measurements of JPR without 
% magnetic field by sweeping frequency will allow one to obtain 
% information on the plasma frequency $ \omega_0(T)\propto 
% 1/\lambda_{ab}^2(T)$ and dissipation of JPR due to quasiparticles (or 
% other degrees of freedom) in high-temperature superconductors and 
% organic superconductors.

\begin{center}
{\bf Acknowledgments}
\end{center}

\begin{figure}
\epsfxsize=3.4in \epsffile{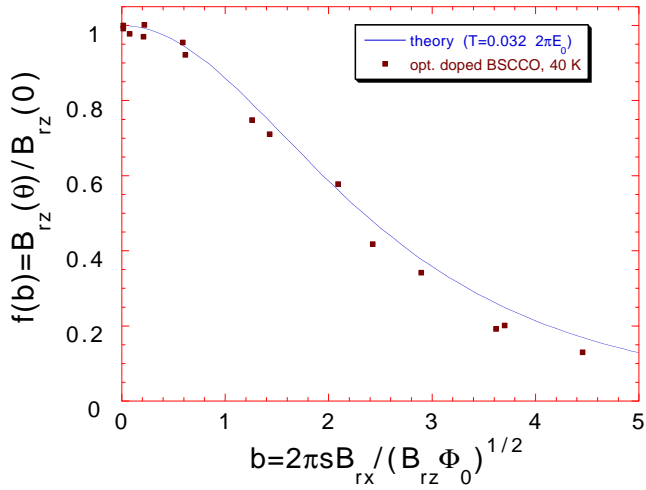} \caption{Comparison between
the function $f(b)=f_{s}(b)/f_{s}(0)$ extracted from the angular
dependence of the resonance field (optimally doped Bi-2212 sample at 40
K from Ref.\ \protect\onlinecite{MatsPRB97}, Fig.\ 3) and numerically
calculated $f(b)$ at $T=0.032\cdot 2\pi E_{0}$}
\label{Fig-fbcomp}
\end{figure}

We thank M.~Gaifullin for providing data of Ref.\ \onlinecite{MatsPRB97} 
which were used in Fig.\ \ref{Fig-fbcomp}.  We would like to thank 
V.~M.~Vinokur for stimulating discussions and I.~Aranson for help in
numerical calculations.  This work in Argonne was supported by the
National Science Foundation Office of the Science and Technology
Center under contract No.  DMR-91-20000.  and by the U.~S.~Department
of Energy, BES-Materials Sciences, under contract No.  W-31-
109-ENG-38.  In Los Alamos this work was supported by U.S. DOE. AEK
also would like to acknowledge support from the Japan Science and
Technology Corporation, STA Fellowship 498051, and to thank National
Research Institute for Metals for hospitality.

\begin{center}
{\bf Appendix: Spectral density in the Lifshitz tail region}
\end{center}

%\section{Appendix: Spectral density in the Lifshitz tail region}

In this Appendix we derive the eigenvalue equation (\ref{DimEq}) which
determines the spectral density $A(E)\propto \exp (-\Phi (E))$ in the
Lifshitz tail region. Using the Lagrange technique we represent $\Phi (E)$
as 
\begin{equation}
\Phi (E)=\min_{u}\left[ {\cal H}\{u_{n}({\bf r})\}-\beta (E-{\cal E}
_{0}\{u\})\right] ,  \label{PhiELag}
\end{equation}
$\beta $ is the Lagrange factor, which has to be determined by the condition 
${\cal E}_{0}\{u\}=E$. Representing ${\cal E}_{0}[u]$ as 
\begin{equation}
{\cal E}_{0}\{u\}=\min_{\Psi }\left[ H_{0}\{\Psi \}+\sum_{n}\int d{\bf r}
u_{n}({\bf r})\Psi _{n}^{2}({\bf r})\right] 
\end{equation}
with the Hamiltonian for the ``free particle'' 
\begin{equation}
H_{0}\{\Psi \}=\lambda _{J}^{2}\sum_{n,m}\int d{\bf r}L_{nm}\nabla \Psi
_{n}\nabla \Psi _{m}
\end{equation}
we can rewrite Eq.\ (\ref{PhiE}) as: 
\begin{equation}
\Phi (E)=-\beta E+\min_{\Psi }\left( \beta H_{0}\{\Psi \}+\min_{u}\left[ 
{\cal H}\{u_{n}({\bf r})\}+\beta \sum_{n}\int d{\bf r}u_{n}({\bf r})\Psi
_{n}^{2}({\bf r})\right] \right) .  \label{phie}
\end{equation}
Optimal $u_{n}({\bf r})$ is determined by the condition 
\begin{equation}
\frac{\delta {\cal H}}{\delta u_{n}({\bf r})}+\beta \Psi _{n}^{2}({\bf r})=0
\end{equation}
which gives 
\begin{equation}
u_{n}({\bf r})=-S_{0}\beta \Psi _{n}^{2}({\bf r})/2,
\end{equation}
and Eq.~(\ref{PhiE}) can be represented as: 
\begin{equation}
\Phi (E)=-\beta E+\min_{\Psi }\left( \beta H_{0}\{\Psi \}-\frac{\beta
^{2}S_{0}}{4}\sum_{n}\int d{\bf r}\Psi _{n}^{4}({\bf r})\right) .
\end{equation}
Optimal $\Psi _{n}({\bf r})$ is determined by the equation 
\begin{equation}
-\lambda _{J}^{2}\sum_{m}L_{nm}\nabla ^{2}\Psi _{m}({\bf r})-(\beta
S_{0}/2)\Psi _{n}^{3}({\bf r})=E\Psi _{n}({\bf r}).
\end{equation}
Applying the operator $-\nabla ^{2}+s^{2}/\lambda _{ab}^{2}$ to this
equation and neglecting small terms of the order $s^{2}/\lambda _{ab}^{2}$
we can rewrite this equation in a simpler form 
\begin{equation}
-\lambda _{J}^{2}\nabla ^{2}\Psi _{n}({\bf r})+(\beta S_{0}/2)\nabla
_{n}^{2}\Psi _{n}^{3}({\bf r})=-E\nabla _{n}^{2}\Psi _{n}({\bf r}),
\end{equation}
where $\nabla _{n}^{2}\Psi _{n}=2\Psi _{n}-\Psi _{n+1}-\Psi _{n-1}$. Using
relation 
\begin{equation}
H_{0}\{\Psi \}=\lambda _{J}^{2}\sum_{n,m}\int d{\bf r}L_{nm}\nabla \Psi
_{n}\nabla \Psi _{m}=\sum_{n}\int d{\bf r}\left( (\beta S_{0}/2)\Psi
_{n}^{4}({\bf r})+E\Psi _{n}^{2}({\bf r})\right) 
\end{equation}
we obtain for $\Phi (E)$: 
\begin{eqnarray}
\Phi (E) &=&-\beta E+\min_{\Psi }\left( \beta H_{0}\{\Psi \}-\frac{\beta
^{2}S_{0}}{4}\sum_{n}\int d{\bf r}\Psi _{n}^{4}({\bf r})\right)  \\
&=&\frac{\beta ^{2}S_{0}}{4}\sum_{n}\int d{\bf r}\Psi _{n}^{4}({\bf r}).
\end{eqnarray}
Introducing dimensionless variables 
%(as in Ref.~\onlinecite{Lifshitz}) 
\begin{equation}
\Psi _{n}({\bf r})=\Psi _{0}(0)\psi _{n}({\bf r}),\ \ \ {\bf r=\tilde{r}}
\frac{\lambda _{J}}{\sqrt{\beta S_{0}\Psi _{0}^{2}(0)/2}},\ \ \ \epsilon
_{0}=\frac{2E}{\beta S_{0}\Psi _{0}^{2}(0)},
\end{equation}
we obtain the dimensionless equation (\ref{DimEq}). $\Phi (E)$ can be found
from solution of this equation as: 
\begin{equation}
\Phi (E)=\frac{\beta }{2}\lambda _{J}^{2}\Psi _{0}^{2}(0)\psi _{4}
\end{equation}
with $\psi _{4}=\sum_{n}\int d{\bf \tilde{r}}\psi _{n}^{4}(\tilde{r})$.
Using the relation $\beta \Psi _{0}^{2}(0)=2E/\epsilon _{0}S_{0}$ we obtain
the final relation (\ref{PhiE1}).


\end{document}